\documentclass[preprint,showpacs,titlepage,aps,prd,
tightenlines,amsmath,byrevtex,nofootinbib]{revtex4}

\usepackage{graphicx}

\def\lsim{\raise0.3ex\hbox{$\;<$\kern-0.75em\raise-1.1ex
\hbox{$\sim\;$}}}
\def\gsim{\raise0.3ex\hbox{$\;>$\kern-0.75em\raise-1.1ex
\hbox{$\sim\;$}}}

\begin{document}

\preprint{hep-ph/0602046}

\title{Recoilless Resonant Absorption of Monochromatic 
Neutrino Beam for Measuring $\Delta m^2_{31}$ and $\theta_{13}$}

%\vskip 2cm

\author{Hisakazu Minakata$^{1, 2}$}
\email{E-mail: minakata@phys.metro-u.ac.jp}
\author{Shoichi Uchinami$^{1}$}
\email{E-mail: uchinami@phys.metro-u.ac.jp}
\affiliation{
$^1$Department of Physics, Tokyo Metropolitan University \\
1-1 Minami-Osawa, Hachioji, Tokyo 192-0397, Japan \\
$^2$Abdus Salam International Center for Theoretical Physics, 
Strada Costiera 11, 34014 Trieste, Italy
}

\date{August 2, 2006}
%\date{\today}

\vglue 1.6cm
%%%%%%%%%%%%%%%%%%%%%%%%%%%%%%%%%%%%%%%%%%%%%%%
%    Abstract
%%%%%%%%%%%%%%%%%%%%%%%%%%%%%%%%%%%%%%%%%%%%%%%

\begin{abstract}

We discuss, in the context of precision measurement of 
$\Delta m^2_{31}$ and $\theta_{13}$, physics capabilities 
enabled by the recoilless resonant absorption of monochromatic 
antineutrino beam enhanced by the M\"ossbauer effect recently 
proposed by Raghavan. 
Under the assumption of small relative systematic error of a few tenth 
of percent level between measurement at different detector locations, 
we give analytical and numerical estimates of the sensitivities to 
$\Delta m^2_{31}$ and $\sin^2 2\theta_{13}$. 
The accuracies of determination of them are enormous; 
The fractional uncertainty in $\Delta m^2_{31}$ achievable 
by 10 point measurement is 0.6\% (2.4\%) for $\sin^2 2\theta_{13} = 0.05$, 
and the uncertainty of $\sin^2 2\theta_{13}$ is 
0.002 (0.008) both at 1$\sigma$ CL with the optimistic (pessimistic) 
assumption of systematic error of 0.2\% (1\%). 
The former opens a new possibility of determining the neutrino mass 
hierarchy by comparing the measured value of $\Delta m^2_{31}$ 
with the one by accelerator experiments,  
while the latter will help resolving the $\theta_{23}$ octant degeneracy. 

\end{abstract}


\maketitle

%%%%%%%%%%%%%%%%%%%%%%%%

\section{Introduction}\label{introduction}

Recently, an intriguing possibility was suggested by Raghavan 
\cite{raghavan1,raghavan2} 
that the resonant absorption reaction \cite{mikaelyan}
\begin{eqnarray}
\bar{\nu_{e}} + ^{3}\text{He} + \text{orbital e}^{-} \rightarrow ^{3}\text{H}
\label{res-abs}
\end{eqnarray}
with simultaneous capture of an atomic orbital electron can be 
dramatically enhanced. 
The key idea is to use monochromatic $\bar{\nu_{e}}$ beam with the 
energy 18.6 keV from the inverse reaction 
$^{3}\text{H}  \rightarrow \bar{\nu_{e}}+$$^{3}\text{He} + \text{orbital e}^{-}$, 
by which the resonance condition is automatically satisfied. 
(See \cite{visscher,schiffer} for earlier suggestions.)  
He then suggested an experiment to 
measure $\theta_{13}$ by utilizing the ultra low-energy monochromatic 
$\bar{\nu_{e}}$ beam. 
Though similar to the reactor $\theta_{13}$ experiments 
\cite{krasnoyarsk,MSYIS,reactor_white}, 
the typical baseline length is of order 10 m due to much lower 
energy of the beam by a factor of $\simeq$150, making it doable 
in the laboratories. 
The mechanism, in principle, would work with more generic setting in which 
$^{3}\text{H}$ and $^{3}\text{He}$ in (\ref{res-abs}) are replaced 
by nuclei A(Z) and A(Z-1).

The author of Ref.~\cite{raghavan1,raghavan2} then went on to 
even more aggressive  proposal of enhancement by a factor of 
$\sim 10^{11}$ by embedding both $^{3}\text{H}$ and $^{3}\text{He}$ 
into solids \cite{visscher,schiffer} 
by which the broadening of the beam due to nuclear recoil 
is severely suppressed by a mechanism similar to the 
M\"ossbauer effect \cite{messbauer}. 
Then, the event rate of the $\theta_{13}$ experiment is enhanced 
by the same factor, allowing extremely high counting rate. 
Thanks to the ultimate energy resolution of 
$\Delta E_{\nu}/E_{\nu} \simeq 2 \times 10^{-17}$ 
enabled by the recoilless mechanism, he was able to propose 
a table top experiment to measure gravitational red shift of neutrinos, 
a neutrino analogue of the Pound-Rebka experiment for photons \cite{pound}.

In this paper, we examine possible physics potentials of the 
$\theta_{13}$ experiment proposed in \cite{raghavan1,raghavan2}. 
The characteristic feature of the experiment, which clearly marks 
the difference from the reactor $\theta_{13}$ measurement, is 
the use of monochromatic beam apart from the shorter baseline 
by a factor of $\simeq$150. 
Then, the most interesting question is how accurately 
$\Delta m^2_{31}$ can be determined. 
Note that even without recoilless setting, the beam energy width is 
of the order of $\Delta E_{\nu}/E_{\nu} \sim 10^{-5}$, and 
it can be ignored for all practical purposes. 
%
%It will allows us to utterly ignore the beam energy width for any thinkable accuracies of measurement of $\Delta m^2_{31}$. 
It is also interesting to explore the accuracy of $\theta_{13}$ measurement. 
In addition to possible extremely high statistics, 
the baseline as short as $\sim$10 m should allow us to utilize the 
setting of continuously movable detector, the one which once 
proposed in a  reactor $\theta_{13}$ experiment \cite{movable} 
but the one that did not survive in the (semi-) final proposal.
The method will greatly help to reduce the experimental systematic 
uncertainties of the measurement.

We will show in our analysis that the accuracies one can achieve for  
$\Delta m^2_{31}$ and $\theta_{13}$ determination by the recoilless 
resonant absorption are enormous. 
At $\sin^2 2\theta_{13} = 0.05$, for example, 
the fractional uncertainty in $\Delta m^2_{31}$ determination is 
0.6\% (2.4\%) and  the uncertainty of $\sin^2 2\theta_{13}$ is 
0.002 (0.008) both at 1$\sigma$ CL under an optimistic (pessimistic) 
assumption of systematic error of 0.2\% (1\%).

What is the scientific merit of such precision measurement of 
$\Delta m^2_{31}$ and $\theta_{13}$?
With a 1\% level precision of $\Delta m^2_{31}$,  
the method for determining neutrino mass hierarchy by comparing 
between the two effective $\Delta m^2$ measured in reactor and 
accelerator (or atmospheric) disappearance measurement 
\cite{NPZ,GJK} would work, opening another door for 
determining the neutrino mass hierarchy. 
It is also proposed \cite{MSYIS} that the $\theta_{23}$ octant degeneracy 
can be resolved by combining reactor measurement of $\theta_{13}$ 
with accelerator disappearance (appearance) measurement of 
$\sin^2 2\theta_{23}$ ($s^2_{23} \sin^2 2\theta_{13}$).\footnote{
%%%%%%%%%%%%%%% footnote %%%%%%%%%%%%%%%%
Here is a short summary for the parameter degeneracy. 
It is the phenomenon that there exist multiple solutions for 
mixing parameters, $\theta_{13}$, $\theta_{23}$ and $\delta$, 
for a given set of measurement of 
$\nu_{\mu}$ disappearance, $\nu_{e}$ and $\bar{\nu}_{e}$ 
appearance probabilities, and the octant ambiguity of 
$\theta_{23}$ is among them.  
The nature of the degeneracy may be characterized as 
the intrinsic degeneracy of $\theta_{13}$ and $\delta$ \cite{intrinsic}, 
which is duplicated by 
the unknown sign of $\Delta m^2_{31}$ \cite{MNjhep01} and 
the octant ambiguity of $\theta_{23}$ for a given $\sin{2\theta_{23}}$ 
\cite{octant}. 
For an overview, see e.g., \cite{BMW,MNP2}.
}
%%%%%%%%%%%%%%% footnote %%%%%%%%%%%%%%%%
%
%
(See \cite{octant,gouvea} for earlier qualitative suggestions.) 
The results of the recent quantitative analysis \cite{resolve23}, 
however, indicate that the resolving power of the method is 
limited at small $\theta_{13}$ primarily because of the uncertainties in 
reactor measurement of $\theta_{13}$. 
Therefore, the highly accurate measurement of 
$\Delta m^2_{31}$ and $\theta_{13}$ 
which is enabled by using the resonant absorption reaction 
should help resolving the mass hierarchy and the 
$\theta_{23}$ degeneracies.

In Sec.~\ref{concept}, we discuss ``conceptual design'' of the 
possible experiments. 
In Sec.~\ref{method}, we define the statistical method for our analysis. 
In Sec.~\ref{numerical}, we present numerical analysis of the sensitivities 
of $\theta_{13}$ and $\Delta m^2_{31}$ measurement. 
In Sec.~\ref{analytic}, we complement the numerical estimate in 
Sec.~\ref{numerical} by giving analytic estimate of the sensitivities. 
In Sec.~\ref{conclusion}, we give some remarks on implications of our results. 
In Appendix~\ref{appendix1}, we give a general formula for the inverse 
of the error matrix.

\section{Which kind of $\theta_{13}$ experiment?}\label{concept}

We give preliminary discussions on which kind of 
setting is likely to be the best one for experiment to measure 
$\Delta m^2_{31}$ and $\theta_{13}$ with use of ultra low-energy 
monochromatic $\bar{\nu_{e}}$ beam. 
In this section, we rely on the one-mass scale dominant \cite{one-mass} 
(or the two-flavor) approximation of the neutrino oscillation probability 
$P(\bar{\nu_{e}} \rightarrow \bar{\nu_{e}})$, 
though we use the full three-flavor expression in our quantitative 
analysis performed in Sec.~\ref{numerical}.  
It reads 
\begin{eqnarray}
P(\bar{\nu_{e}} \rightarrow \bar{\nu_{e}}) = 1 - 
\sin^2 2\theta_{13} 
\sin^2 {\left(
\frac{\Delta m^2_{31} L}{4 E}
\right)}, 
\label{Pee}
\end{eqnarray}
where the neutrino mass squared difference is defined as 
$\Delta m^2_{ji} \equiv m^2_{j} -  m^2_{i}$ with neutrino masses 
$m_{i}$ ($i$ = 1 - 3)\footnote{
%%%%%%%%%%%%% footnote %%%%%%%%%%%%%%%%%
When we speak about discriminating the neutrino mass hierarchy 
by comparing the two ``large'' $\Delta m^2$ measured in 
$\bar{\nu}_{e}$ disappearance and 
$\nu_{\mu}$ disappearance channels, one has to be careful about  the 
definition of $\Delta m^2$ which enter into the survival probabilities \cite{NPZ}.
While keeping this point in mind, we do not try to elaborate the 
expressions of $\Delta m^2$ in this paper by just writing it as 
$\Delta m^2_{31}$ in $\bar{\nu}_{e}$ 
disappearance channel which may be interpreted as 
$\Delta m^2_{\text{eff}}|_{e}$ in \cite{NPZ}. 
}
and $L$ is a distance from a source to a detector.  
With $E_{\nu} = 18.6$ keV, the first oscillation maximum 
(minimum in $P(\bar{\nu_{e}} \rightarrow \bar{\nu_{e}})$) 
is reached at the baseline distance 
\begin{eqnarray}
L_{\text{\text{OM}} } = 9.2 
\left(  \frac{\Delta m^2_{31}}{2.5 \times 10^{-3} \text{eV}^2}  \right)^{-1} 
\text{m}. 
\label{LOM}
\end{eqnarray}
While the current value of $\Delta m^2_{31}$ which comes from 
the atmospheric \cite{SKatm} and the accelerator \cite{K2K} 
measurement has large uncertainties, it should be possible to 
narrow down the value thanks to the ongoing and the forthcoming 
disappearance measurement by MINOS 
\cite{MINOS} and T2K \cite{JPARC} experiments. 
Furthermore, the experiment considered in this paper is powerful 
enough to determine both quantities accurately at the same time, 
if detector locations are appropriately chosen.

The whole discussion of the $\theta_{13}$ experiment 
must be preceded by the test measurement 
at $\sim$10 cm or so to verify the principle, namely to 
demonstrate that the mechanism of resonant enhancement 
proposed in \cite{raghavan1,raghavan2} is indeed at work. 
At the same time, the flux times cross section must be 
measured to check the consistency of the Monte Carlo estimate. 
Then, one can go on to the measurement of $\theta_{13}$ and 
$\Delta m^2_{31}$, and possibly other quantities. 
Because of the expected high statistics of the experiment it is 
natural to think about using spectrum informations. 
In the case of monochromatic beam it amounts to consider 
measurement at several different detector locations.

Let us estimate the event rate. 
Although the precise rate is hard to estimate, 
the numbers displayed below will give the readers a feeling 
on what would be the time scale for the experiment. 
The $\bar{\nu_{e}}$ flux from $^3$H source with strength $S$ MCi 
due to bound state beta decay is given by 
\begin{eqnarray}
f_{\bar{\nu_{e}}} = 1.4 \times 10^{7}
\left(  \frac{S}{1 \text{MCi} }  \right) 
\left(  \frac{L}{10~\text{m} }  \right)^{-2} 
\text{cm}^{-2} \text{s}^{-1}
\label{flux}
\end{eqnarray}
where the ratio of bound state beta decay to free space decay 
is taken to be $4.7 \times 10^{-3}$  based on 
\cite{bahcall}.
The rate of the resonant absorption reaction can be computed 
by using cross section $\sigma_{res}$ and number of target atoms 
$N_{T}$ as 
$R = N_{T} f_{\bar{\nu_{e}}}  \sigma_{res}$. 
Without the M\"ossbauer enhancement the cross section is estimated to be 
$\sigma_{res} \simeq 10^{-42} \text{cm}^2$ \cite{raghavan1,raghavan2} 
based on \cite {mikaelyan}. 
Then, the rate with target mass $M_{T}$ without neutrino oscillation 
is given by 
\begin{eqnarray}
R = 2.4 \times 10^{-4}
\left(  \frac{S M_{T}}{1 \text{MCi} \cdot \text{kg} }  \right) 
\left(  \frac{L}{10~\text{m} }  \right)^{-2} 
\text{day}^{-1} 
\label{rate}
\end{eqnarray}
An improved 
estimate in \cite{raghavan2} entailed a factor of 
$\simeq 10^{11}$ enhancement of the cross section 
by the M\"ossbauer effect after the source and the target are 
embedded into solids.  Assuming the enhancement factor, 
$\sigma_{res} \simeq 5 \times 10^{-32} \text{cm}^2$ and 
the rate becomes 
\begin{eqnarray}
R_{\text{enhanced}} = 1.2 \times 10^{4}
\left(  \frac{S M_{T}}{1 \text{MCi} \cdot \text{g} }  \right) 
\left(  \frac{L}{10~\text{m} }  \right)^{-2} 
\text{day}^{-1} 
\label{enhancedR}
\end{eqnarray}
Therefore, one obtains about $1.2 \times 10^{6}$ events per day  
for 1 MCi source and 100 g $^3$He target at a baseline distance $L=10$ m.
If the enhancement factor is not reached the running time for 
collecting the same number of events becomes longer accordingly.

Thus, once the $^3$He (and much easier $^3$H) implementation 
into solid is achieved, the event rate is sufficient. 
The real issue for high sensitivity measurement of 
$\Delta m^2_{31}$ and $\sin^2 2\theta_{13}$  is whether the produced 
$^3$H can be counted directly without waiting for decaying back to 
$^3$He by emitting electron. 
It is because the long lifetime of 12.33 year \cite{table_isotope} 
of $^3$H makes it impossible to identify which period the decayed 
$^3$H was produced, resulting in the errors of the event rate in 
each detector location. 
Possibilities of real-time counting and direct counting by extracting 
$^3$H atoms are mentioned in \cite{raghavan2}. 
In this paper, we assume that at least one of such methods works, 
and it offers opportunity of direct counting of events. 
Note that the detection efficiency need not to be high because of 
huge number of events. What is important is the time-stable 
counting rate which allows relative systematic errors between 
measurement at different detector locations small enough.

\section{Statistical method for analysis}\label{method}

In this section, we define the statistical procedure for our analysis to 
estimate the sensitivities of $\Delta m^2_{31}$ and $\sin^2 2\theta_{13}$ 
to be carried out in the following sections. 
We aim at illuminating general properties of the 
$\chi^2$ under the assumption of the small uncorrelated systematic 
errors compared to the correlated ones.

\subsection{Definition of $\chi^2$ and characteristic properties of errors}

We consider measurement at $n$ different distances 
$L=L_{i}$ ($i = 1, 2, ... n$) from the source. 
Then, the appropriate form of $\Delta\chi^2$ which is suited for 
analytic study \cite{SYSHS} and is simply denoted as $\chi^2$ hereafter, 
is as follows: 
\begin{eqnarray}
\chi^2 = 
\sum_{i =1}^{n}
\frac{
\left[ N^{obs}_{i} - (1+\alpha) N^{exp}_{i} \right]^2}{N^{exp}_{i} + (\sigma_{usys,i} N^{exp}_{i})^2}  + 
\left( \frac{\alpha}{\sigma_{c}} \right)^2 
\label{chi2_1}
\end{eqnarray}
where $N^{obs}$ is the number of events computed with the values 
of parameters given by nature, and $N^{exp}$ is the one computed with 
certain trial set of parameters. 
$\sigma_{c}$ is the systematic error common to measurement 
at $n$ different distances, the correlated error, whereas 
$\sigma_{usys,i}$ indicate errors that cannot be attributed to 
$\sigma_{c}$, the uncorrelated errors.
The example of the former and the latter errors are as follows:

\begin{itemize}

\item

$\sigma_{c}$ (correlated error):
Uncertainties in number of target $^3$He atoms, 
errors in counting the number of produced tritium nuclei, 
errors in calculating resonant absorption cross section, 
errors in estimating the efficiency of counting tritium nuclei 

\item

$\sigma_{usys,i}$ (uncorrelated error) :
Possible time dependences of number of decaying tritium nuclei 
and detection efficiency of events

\end{itemize}

Since we consider moving detector setting the list of the thinkable 
uncorrelated systematic errors is quite limited. 
If the near detector with the identical structure with a movable 
far detector exists the error can, in principle, be vanishingly small. 
One may think of the errors of the order of 0.1\%-0.3\%. 
It is because the flux times cross section can be monitored in real time 
by a near detector. 
In fact, the similar values for uncorrelated systematic error 
are adopted in sensitivity estimate of some of the reactor $\theta_{13}$ 
experiments such as the Braidwood, the Daya Bay, 
and the Angra projects \cite{braidwood,dayabay,angra}. 
In near future experiments, somewhat larger values are taken, 
0.6\% in Double-Chooz project \cite{DCHOOZ} and 
0.35\% in KASKA \cite{KASKA}.

On the other hand, it may not be so easy to control the correlated 
systematic error $\sigma_{c}$. 
The number of $^3$H nuclei may be measured when they are 
implemented into solid. 
The number of target nuclei times the resonant absorption cross 
section may be measured in a research and development stage 
with a near detector. 
Therefore, we suspect that the largest error may come from 
uncertainty in counting rate of the produced $^3$H nuclei.
Of course, reliable estimate of systematic errors $\sigma_{c}$ and 
$\sigma_{usys}$ requires specification 
of the site to estimate the background caused by n$^3$H reaction etc. 
But, it can be experimentally measured by the source on and off procedure,
as pointed out in \cite{raghavan1}.
Lacking definitive numbers for $\sigma_{c}$ at the moment, 
we use a tentative value $\sigma_{c} = 10$\% throughout our analysis.
We have checked that the results barely change even if we use 
more conservative number $\sigma_{c} = 20$\%. 

%??? IF DIRECT COUNTING DOES NOT WORK ???

If the direct counting of $^3$H atoms does not work, we may have 
to expect much larger systematic errors, because one has to 
extract event rate at each detector location only by fitting the decay curve. 
In this case, determination of baseline dependent event rates would 
be more and more difficult for larger number of detector locations. 
Probably, the better strategy without the direct counting would be to 
place multiple identical detectors (or of the same structure) at 
appropriate baseline distances. 
Even in this case, it is quite possible that the uncorrelated 
systematic error $\sigma_{usys}$ can be controlled to 1\% level, 
as expected in a variety of reactor $\theta_{13}$ experiments 
\cite{reactor_white}. 

\subsection{Approximate form of $\chi^2$ with hierarchy in errors}\label{hierarchy}

By eliminating $\alpha$ through minimization the $\chi^2$ can be written as 
\begin{eqnarray}
\Delta\chi^2= 
\vec{x}^{\text{T}} V^{-1} \vec{x} 
\label{chi2_2}
\end{eqnarray}
where $\vec{x}$ is defined as  
\begin{eqnarray}
\vec{x}^{\text{T}} = 
\left[
\frac{N^{obs}_{1} - N^{exp}_{1}}{N^{exp}_{1}}, 
\frac{N^{obs}_{2} - N^{exp}_{2}}{N^{exp}_{2}}, \cdot  \cdot \cdot 
\frac{N^{obs}_{n} - N^{exp}_{n}}{N^{exp}_{n}}
\right]. 
\label{vector}
\end{eqnarray}
Using the general formula given in Appendix~\ref{appendix1}, $V^{-1}$ is given by 
\begin{eqnarray}
(V^{-1})_{i j} =
\frac{\delta_{i j}}{ \sigma^2_{u i}  } -
\frac{ 1 }
{\sigma^2_{u i}  \sigma^2_{u j} }
\frac{ \sigma^2_{c} }
{ \left[
1 + \left(
\sum_{k=1}^{n} \frac{1}{\sigma^2_{u k}}
\right) 
\sigma^2_{c} 
\right] }
\label{Vinv}
\end{eqnarray}
where 
$\sigma^2_{u i} \equiv \sigma^2_{usys,i} + \frac{1}{N^{exp}_{i}}$.
%This formula will be used in our analytic estimate of the sensitivities in Sec.~\ref{analytic}. 
By construction, the $\chi^2$ depends upon $\sigma_{usys,i}$ and 
$N^{exp}_{i}$ only through this combination. 
%$\sigma^2_{u i} \equiv \sigma^2_{usys,i} + \frac{1}{N^{exp}_{i}}$. 
Therefore, the particular case that will be taken in the next section, in fact, 
includes many cases with different event number but with the 
same $\sigma_{u}$.

Under the approximation $\sigma^2_{ui} \ll \sigma^2_{c}$, 
$V^{-1}$ simplifies;
\begin{eqnarray}
(V^{-1})_{i j} =
\frac{\delta_{i j}}{ \sigma^2_{u i}  } -
%\frac{ 1 }{\sigma^2_{u i}  \sigma^2_{u j} }
%
\frac{1 }
{ \sigma^2_{u i}  \sigma^2_{u j} \left( \sum_{k=1}^{n} \frac{1}{\sigma^2_{u k}} \right)  }.
\label{Vinv_corrdom}
\end{eqnarray}
The remarkable feature of (\ref{Vinv_corrdom}) is the ``scaling behavior'' 
in which $\chi^2$ is independent of the correlated error $\sigma_{c}$, 
and the sensitivity to $\sin^2\theta_{13}$ and 
$\Delta m^2_{31}$ can be made higher as the uncorrelated systematic errors 
as well as the statistical error become smaller. 
It may be counterintuitive because the leading term of the 
error matrix $V$ is of order $\sigma^2_{c}$. 
(See Appendix \ref{appendix1}.)
It is due to the singular nature of the leading order matrix, 
as noted in \cite{reactorCP}.

%\newpage

\section{Estimation of sensitivities of $\Delta m^2_{31}$ and 
$\theta_{13}$}\label{numerical}

We now examine the sensitivities of $\Delta m^2_{31}$ and 
$\sin^2 2\theta_{13}$ achievable by the recoilless resonant absorption 
of monochromatic $\bar{\nu}_{e}$ enhanced by the M\"ossbauer effect. 
The numerical estimate of the sensitivities in this section 
will be followed by the one by the analytic method in Sec.~\ref{analytic}.

The setting of movable detector and the expected high statistics of the 
experiment make it possible to consider the situation that an 
equal number of events are taken in each detector location. 
Of course, the far a detector from the source, the longer an exposure will take. 
In the following analysis, the number of events are assumed to be 
$10^{6}$ in each detector location. 
Given the rate in (\ref{enhancedR}), and assuming that the direct 
counting works, 
it is obtainable in $\sim$10 days for 100 g $^3$He target 
even if the detector is located at the second oscillation maximum, 
$L=3 L_{\text{OM}} $. 
On the other hand, the number of events $10^{6}$ is sufficient 
for our purpose because it is unlikely that the uncorrelated 
systematic errors can be made much smaller than 0.1\%.

We take a ``common-sense approach'' to determine 
the locations of the detectors and postpone the discussion of the 
optimization problem. 
We examine the following four types of run, Run I, IIA, IIB, and III, 
for the measurement. 

\begin{itemize}

\item

Run I: 
Measurement at 5 detector positions, 
$L=
\frac{1}{5} L_{\text{OM}} , 
\frac{3}{5} L_{\text{OM}} , 
L_{\text{OM}} , 
\frac{7}{5} L_{\text{OM}} $, and $\frac{9}{5} L_{\text{OM}} $
are considered so that 
$\Delta \equiv \Delta m^2_{31} L / 4 E = 
\pi/10, 3\pi/10, \pi/2,  7\pi/10,  9\pi/10$ are covered. 

\item

Run IIA: 
A setting for precision determination of $\Delta m^2_{31}$ by 
measurement at 10 detector positions: 
$L= L_{i}$ ($i=1, ... 10$) where $L_{i+1} = L_{i} + \frac{1}{5} L_{\text{OM}} $ and 
$L_{1} = \frac{1}{5} L_{\text{OM}} $  
so that the range $\Delta = 0$ to $\pi$ is covered.

\item

Run IIB: 
A setting for precision determination of $\Delta m^2_{31}$ by 
measurement at 10 detector positions: 
$L= L_{i}$ ($i=1, ... 10$) where $L_{i+1} = L_{i} + \frac{2}{5} L_{\text{OM}} $ and 
$L_{1} = \frac{1}{5} L_{\text{OM}} $  
so that the entire period, 
$\Delta = 0$ to 2$\pi$, is covered. 

\item

Run III: 
A setting of 20 detector positions: 
$L= L_{i}$ ($i=1, ... 20$) where 
$L_{i+1} = L_{i} + \frac{1}{5} L_{\text{OM}} $ and $L_{1} = \frac{1}{5} L_{\text{OM}} $ 
($\Delta = 0 - 2\pi$). 
It is to check the scaling behavior of the sensitivity with respect to errors.

\end{itemize}

%% NOTE HERE ON THE SCALING.N=million, $\sigma_{usys}=0.2$\% is equivalent to N=200,000 and $\sigma_{usys}=0$

In the following two subsections \ref{optimistic} and \ref{pessimistic}, 
we examine the cases of 
the optimistic ($\sigma_{usys}=0.2$\%) and 
the pessimistic ($\sigma_{usys}=1$\%) systematic errors. 
We stress here that the analyses we will present there contain 
much more general cases. 
%\footnote{
%%%%%%%%%%%%%%% footnote %%%%%%%%%%%%%%%%
For example, because of the scaling behavior 
discussed in the previous section, the case with 
$N=10^{6}$ and $\sigma_{usys}=0.2$\%  is equivalent to 
$N= 2 \times 10^{5}$ and $\sigma_{usys}=0.0$\%. 
Similarly, the case with 
$N=10^{6}$ and $\sigma_{usys}=1$\%  is equivalent to 
$N= 1.33 \times 10^{4}$ and $\sigma_{usys}=0.5$\%.
%}
%
In the last subsection \ref{indirect}, we give an estimate of 
the sensitivities using a tentative setting which may be possible 
without direct counting of $^3$H atoms.

\subsection{Case of optimistic systematic error} 
\label{optimistic}

We focus in this subsection on the case of optimistic systematic 
error, from which one may obtain some feeling on the ultimate 
sensitivities achievable by the present method with the four Run options 
described above. 
As we mentioned earlier, the correlated systematic error 
$\sigma_{c}$ is  taken to be a tentative value of 10\% 
throughout our analysis. 
%Stability of the results will be checked by examining the cases $\sigma_{c}=10\%$ and 2\%.  
The uncorrelated systematic error $\sigma_{usys}$, 
which is assumed to be equal for all detector locations, 
is taken to be 0.2\% in this subsection.

%%%%%%%%%%%% FIGURE I %%%%%%%%%%%%%%%
\begin{figure}[htbp]
%\vglue -0.5cm
\begin{center}
\includegraphics[width=0.76\textwidth]{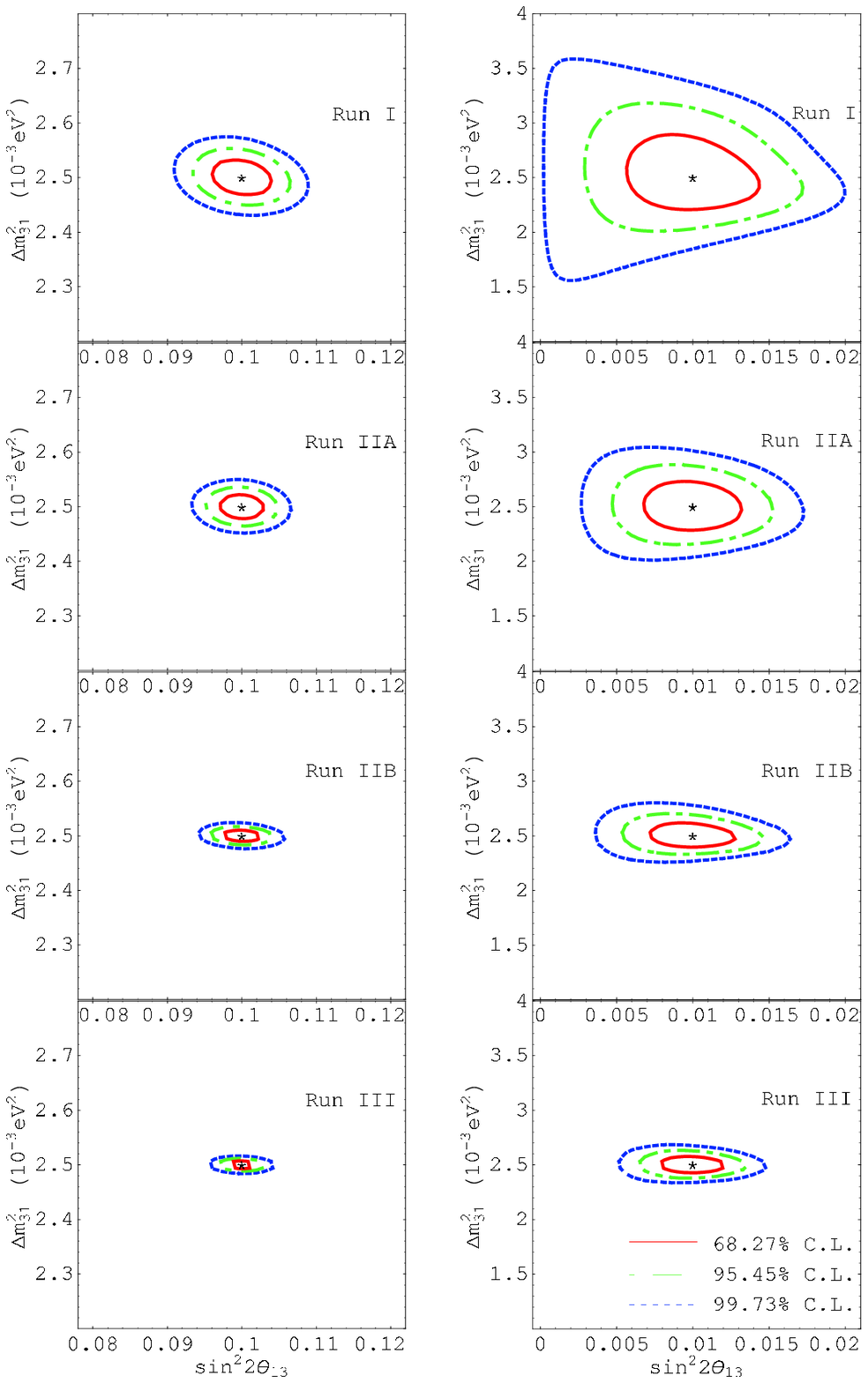}
\end{center}
\vglue -0.2cm
\caption{
The expected allowed region by Run I, IIA, IIB, and III with 
number of events $10^{6}$ in each location are depicted. 
The red-solid, the green-dashed, and the 
blue-dotted lines are for 1$\sigma$ (68.27\%),  
2$\sigma$ (95.45\%), and 3$\sigma$ (99.73\%) CL for 2 DOF, respectively. 
The input values of the mixing parameters are marked by asterisks and 
they are as follows: 
$\Delta m^2_{31} = 2.5 \times 10^{-3}$ eV$^2$, 
$\sin^2 2\theta_{13}=0.1$ and 0.01 
in the left and the right panels. 
}
\label{2d_contour_opt}
\end{figure}
%%%%%%%%%%%% FIGURE I %%%%%%%%%%%%%%%

%%%%%%%%%%%%%%%%%%%% Table I %%%%%%%%%%%%%%%%%%%%%%
\begin{table}
\caption[aaa]{
The expected fractional uncertainty $\delta(\Delta m^2) / \Delta m^2_{31}(0)$ in \% 
for the optimistic systematic error of $\sigma_{usys} = 0.2$\% reachable by the 
Run I$-$III defined in the text. 
The uncertainties are given at 1$\sigma$ (68.27\%) CL for 1 DOF, 
and the numbers in parentheses are the ones at 
3$\sigma$ (99.73\%) CL for 1 DOF. 
In the left, middle, and right columns, the input value of $\theta_{13}$ 
are taken as 
$\sin^2 2\theta_{13}=0.1$, 0.05, and 0.01, respectively.   
}
\vglue 0.5cm
\begin{tabular}{c|ccc}
%\hline
%\multicolumn{3}{c}{54 km}\\
\hline
\ $\sigma_{usys}=0.2$\% \  &\ \ $\sin^2 2\theta_{13}=0.1$ \
             &\ \ $\sin^2 2\theta_{13}=0.05$ \
             &\ \ $\sin^2 2\theta_{13}=0.01$ \ \\
             \hline
\ Run type \  &\ \  \
             & $\delta(\Delta m^2) / \Delta m^2_{31}(0)$ (in \%)  at 1$\sigma$ (3$\sigma$) CL  
             &\ \  \\
\hline
Run I (5 locations)
                 &\ \ 0.84 (2.5) 
                 &\ \ 1.7 (5.0)
                 &\ \ 9.6 $\left( ^{+31}_{-23} \right)$ \\
\hline
Run IIA (10 locations)
                 &\ \ 0.56 (1.7) 
                 &\ \ 1.2 (3.5)
                 &\ \ 6.0 $\left( ^{+18}_{-16} \right)$ \\
\hline
Run IIB (10 locations)
                 &\ \ 0.28 (0.8) 
                 &\ \ 0.56 (1.6)
                 &\ \ 2.8 $\left( ^{+9.6}_{-8.0} \right)$ \\
\hline
Run III (20 locations)
                 &\ \ 0.2 (0.56) 
                 &\ \ 0.4 (1.2)
                 &\ \ 2.0 $\left( ^{+6.0}_{-5.6} \right)$ \\
\hline
\hline
\end{tabular}
\label{Tabmass}
%\vglue 0.5cm
\end{table}
%%%%%%%%%%%%%%%%%%%% Table I %%%%%%%%%%%%%%%%%%%%%%
%
%
%%%%%%%%%%%%%%%%%%%% Table II %%%%%%%%%%%%%%%%%%%%%%
\begin{table}
\caption[aaa]{
The expected sensitivity to the $\sin^2 2\theta_{13}$ for the optimistic 
systematic error of $\sigma_{usys} = 0.2$\% reachable by the 
Run I$-$III defined in the text. 
The uncertainties $\delta(\sin^2 2\theta_{13})$ 
are given at 1$\sigma$ (68.27\%) CL for 1 DOF, 
and the numbers in parentheses are the ones at 
3$\sigma$ (99.73\%) CL for 1 DOF. 
In the left, middle, and right columns, the input value of $\theta_{13}$ 
are taken as 
$\sin^2 2\theta_{13}=0.1$, 0.05, and 0.01, respectively. 
}
\vglue 0.5cm
\begin{tabular}{c|ccc}
%\hline
%\multicolumn{3}{c}{54 km}\\
\hline
\ \ $\sigma_{usys}=0.2$\%  \  &\ \ $\sin^2 2\theta_{13}=0.1$ \
             &\ \ $\sin^2 2\theta_{13}=0.05$ \
             &\ \ $\sin^2 2\theta_{13}=0.01$ \\
             \hline
\ Run type \  &\ \  \
             &\ \ ~~$\sin^2 2\theta_{13}$ at 1$\sigma$ (3$\sigma$) CL~~ \
             &\ \  \ \\
\hline
Run I (5 locations)
                 &\ \ 0.1 $\pm 0.0026$ (0.0078) 
                 &\ \ 0.05 $\pm 0.0027$ (0.0081)
                 &\ \ 0.01 $\pm 0.0028$ (0.0085)  \\
\hline
Run IIA (10 locations)
                 &\ \ 0.1 $\pm 0.0019$ (0.0058) 
                 &\ \ 0.05 $\pm 0.0020$ (0.0061)
                 &\ \ 0.01 $\pm 0.0021$ (0.0064)  \\
\hline
Run IIB (10 locations)
                 &\ \ 0.1 $\pm 0.0017$ (0.0050)  
                 &\ \ 0.05 $\pm 0.0018$ (0.0053)
                 &\ \ 0.01 $\pm 0.0018$ (0.0055)  \\
\hline
Run III (20 locations)
                 &\ \ 0.1 $\pm 0.0013$ (0.0038) 
                 &\ \ 0.05 $\pm 0.0014$ (0.0041)
                 &\ \ 0.01 $\pm 0.0014$ (0.0042)  \\
\hline
\hline
\end{tabular}
\label{Tabangle}
\vglue 0.5cm
\end{table}

%%%%%%%%%%%%%%%%%%%%% Table II %%%%%%%%%%%%%%%%%%%%%%

In Fig.~\ref{2d_contour_opt} we show in 
$\sin^2 2\theta_{13} - \Delta m^2_{31}$ plane 
the expected allowed region by Run I, IIA, IIB, and III with 
number of events $10^{6}$ in each location. 
Throughout the analysis, the true values of $\Delta m^2_{31}$ 
is assumed to be $\Delta m^2_{31} = 2.5 \times 10^{-3}$ eV$^2$. 
The input values of $\sin^2 2\theta_{13}$ is taken as 0.1 and 0.01 
in the left and the right panels in Fig.~\ref{2d_contour_opt}, respectively. 
Throughout the numerical analyses in this paper, 
the other oscillation parameters are taken as: 
$\Delta m^2_{21} = 7.9 \times 10^{-5}$ eV$^2$, 
$\sin^2 \theta_{12} = 0.31$, and $\sin^2 \theta_{23} = 0.5$. 
%%%%%%%%%%% CONFIRMED %%%%%%%%%%%%%
%
In each panel, the red-solid, the green-dashed, and the 
blue-dotted lines are for 1$\sigma$ (68.27\%),  
2$\sigma$ (95.45\%), and 3$\sigma$ (99.73\%) CL for 2 DOF 
(degrees of freedom), respectively.

To complement Fig.~\ref{2d_contour_opt}, 
we give in Table~\ref{Tabmass} the expected sensitivities to 
$\Delta m^2_{31}$ at 1$\sigma$ and 3$\sigma$ CL 
(the latter in parentheses) for 1 DOF for Run I-III. 
They are obtained by optimizing $\sin^2 2\theta_{13}$ in the fit. 
For relatively large $\theta_{13}$, 
$\sin^2 2\theta_{13} \gsim 0.05$ the expected sensitivities to 
$\Delta m^2_{31}$ are enormous. 
For $\sin^2 2\theta_{13} = 0.05$ the sensitivities are already less 
than 1\% in Run IIA, and is about 0.6\% in Run IIB both at 1$\sigma$ CL.
 The scaling behavior mentioned at the end of the previous section 
 is roughly satisfied, as indicated in Table~\ref{Tabmass}. 
(See Sec.~\ref{analytic} for more detailed discussions.) 
For a small value of $\theta_{13}$, $\sin^2 2\theta_{13} = 0.01$
the sensitivity to $\Delta m^2_{31}$ are much worse, as shown in 
Table~\ref{Tabmass}. 
They are about 6\% in Run IIA, and 3\% in Run IIB both at 1$\sigma$ CL.
If Run III is carried out it can go down to 2\%.

In Table~\ref{Tabangle} the expected sensitivity to  
$\sin^2 2\theta_{13}$ at 1$\sigma$ and 3$\sigma$ CL 
(the latter in parentheses) for 1 DOF are given. 
The sensitivities to $\sin^2 2\theta_{13}$ can be better 
characterized by $\delta(\sin^2 2\theta_{13})$, not its fraction to 
$\sin^2 2\theta_{13}$, as will be understood in our analytic treatment 
in Sec.~\ref{analytic}. 
By Run I one can already achieve the accuracy of 
$\delta(\sin^2 2\theta_{13}) \simeq 0.003$, and Run IIA or IIB reach to 
$\delta(\sin^2 2\theta_{13}) \simeq 0.002$. 
The effect of measurement at multiple detector locations on 
improvement of the sensitivity is relatively minor 
in the case of sensitivities to $\sin^2 2\theta_{13}$. 
This is in sharp contrast to the case of $\Delta m^2_{31}$.

%%%%%%%%%%%% FIGURE II %%%%%%%%%%%%%%%
\begin{figure}[htbp]
\vglue 0.5cm
\begin{center}
\includegraphics[width=0.64\textwidth]{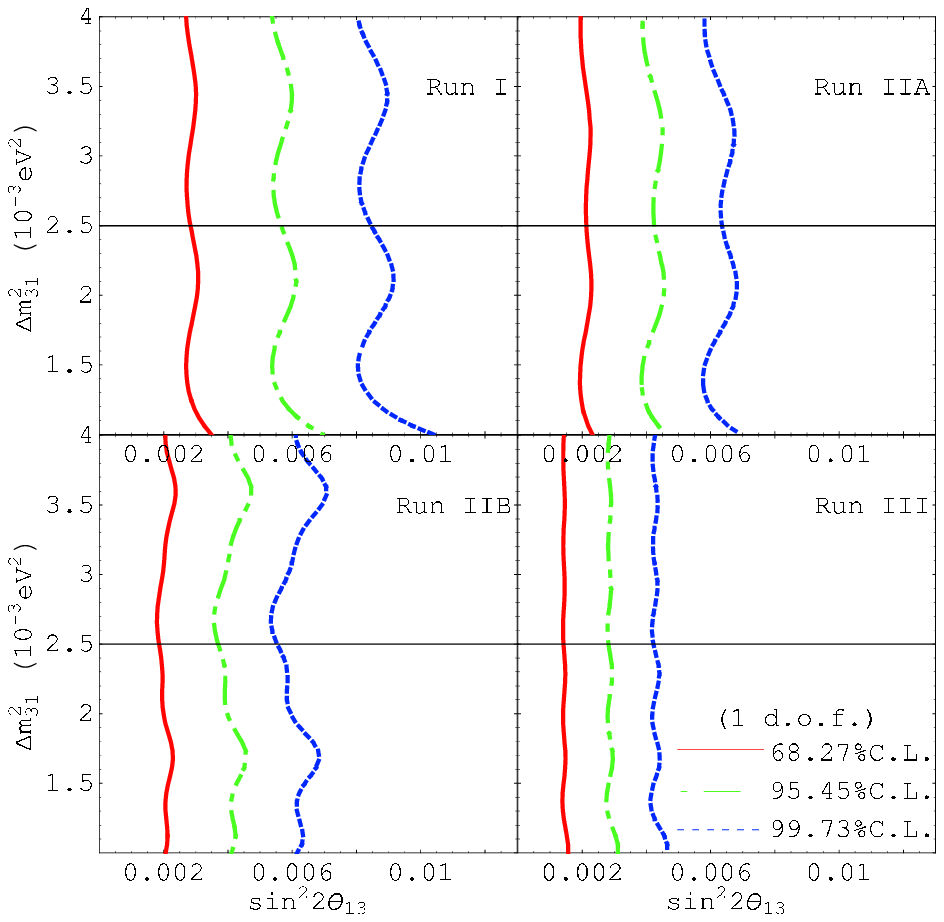}
\end{center}
%\vglue -0.5cm
\caption{
The sensitivity limit of $\theta_{13}$ by Run I, IIA, IIB, and III with 
number of events $10^{6}$ in each location are depicted. 
The red-solid, the green-dashed, and the 
blue-dotted lines are for 1$\sigma$ (68.27\%),  
2$\sigma$ (95.45\%), and 3$\sigma$ (99.73\%) CL for 1 DOF, 
respectively. 
}
\label{1d_exclude}
\end{figure}
%%%%%%%%%%%% FIGURE II %%%%%%%%%%%%%%%

To show the sensitivity limit on $\theta_{13}$ achievable by the 
present method, we present in Fig.~\ref{1d_exclude} the 
excluded regions in $\sin^2 2\theta_{13} - \Delta m^2_{31}$ 
space, assuming the case of no depletion of $\bar{\nu}_{e}$ flux.
The four panels in Fig.~\ref{1d_exclude} correspond to 
Run I, IIA, IIB, and III. 
In each panel, the red-solid, the green-dashed, and the 
blue-dotted lines are for 1$\sigma$ (68.27\%),  
2$\sigma$ (95.45\%), and 3$\sigma$ (99.73\%) CL for 1 DOF, 
respectively. 
The sensitivities indicated in Fig.~\ref{1d_exclude} is quite 
impressive, which reach to 
$\sin^2 2\theta_{13} \simeq 0.006$ at 2$\sigma$ CL even in Run I, and to 
$\sin^2 2\theta_{13} \simeq 0.004$ at the same CL in Run IIB. 
As expected the improvement by adding more 
detector locations is relatively minor.

\subsection{Case of pessimistic systematic error} 
\label{pessimistic}

\begin{figure}[htbp]
%\vglue -0.5cm
\begin{center}
\includegraphics[width=0.76\textwidth]{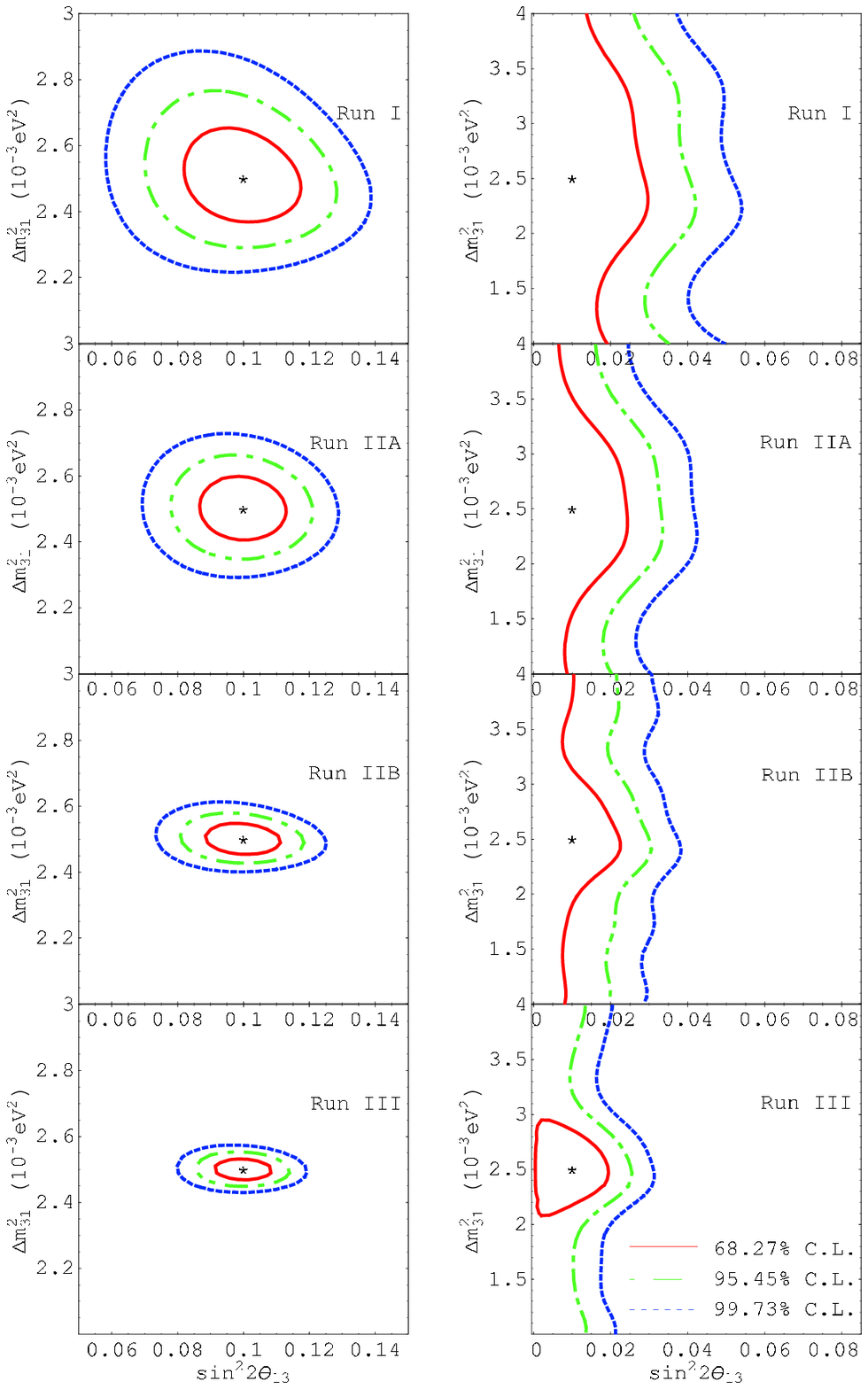}
\end{center}
\vglue -0.2cm
\caption{
The same as in Fig.~\ref{2d_contour_opt} but with the pessimistic systematic error of $\sigma_{usys}$ 1\%.The input values of $\sin^2 2\theta_{13}$ is taken as 0.1 and 0.01 in the left and the right panels. 
}
\label{2d_contour_pessi}
\end{figure}

It might be possible that we end up with the error of $\sim$1\% 
due to, e. g., time dependence of the source even though the method 
of movable detector with direct counting of $^3$H works. 
In Fig.~\ref{2d_contour_pessi} we present the similar allowed region in 
$\sin^2 2\theta_{13} - \Delta m^2_{31}$ space obtained by the same 
Run I, IIA, IIB, and III with the same number of events of 
$10^{6}$ in each location but with 
a pessimistic systematic error of $\sigma_{usys}=$ 1\%. 
At large $\theta_{13}$, $\sin^2 2\theta_{13}=0.1$, we still have 
reasonable sensitivities to $\Delta m^2_{31}$. 
For Run IIB and III, for example, the sensitivities are about 1-2\% level 
for 2 DOF. 
At $\sin^2 2\theta_{13}=0.01$, however, the sensitivity to 
$\Delta m^2_{31}$ is lost except for the one at 1$\sigma$ CL in Run III.
It indicates that the value of $\theta_{13}$ is close to the sensitivity 
limit, and hence we do not place the figure for it though we did for the 
case of optimistic error, Fig.~\ref{1d_exclude}.

%%%%%%%%%%%% FIGURE III %%%%%%%%%%%%%%%
%%%%%%%%%%%% FIGURE III %%%%%%%%%%%%%%%

For more detailed information on sensitivities with the 
pessimistic systematic error of $\sigma_{usys} = 1$\%, we give 
in Tables~\ref{Tabmass_pessi} and \ref{Tabangle_pessi} 
the  sensitivities at 1$\sigma$ and 3$\sigma$ CL (the latter in parentheses) 
for 1 DOF to $\Delta m^2_{31}$ and $\sin^2 2\theta_{13}$, respectively. 
The column without number represents that no limit is obtained, 
in the similar way as seen in the right panels in 
Fig.~\ref{2d_contour_pessi}. 
At relatively large $\theta_{13}$, 
$\sin^2 2\theta_{13}=0.1$ and 0.05, the sensitivities to 
$\Delta m^2_{31}$ remain to be good, 1.2\% and 2.4\% 
at 1$\sigma$ CL for Run IIB. 
But,  the sensitivity quickly drops and is 
about 15\% in the same Run for $\sin^2 2\theta_{13}=0.01$.
The sensitivity to $\sin^2 2\theta_{13}$ is still reasonable, 
$\delta(\sin^2 2\theta_{13}) \simeq 0.008$ at 1$\sigma$ CL for Run IIB 
even at $\sin^2 2\theta_{13}=0.01$.

%%%%%%%%%%%%%%%%%%%% Table III %%%%%%%%%%%%%%%%%%%%%%
\begin{table}
\caption[aaa]{
The expected fractional uncertainty $\delta(\Delta m^2) / \Delta m^2_{31}(0)$ in \% 
for the pessimistic  systematic error of $\sigma_{usys} = 1$\% reachable by the 
Run I$-$III defined in the text. 
The uncertainties are given at 1$\sigma$ (68.27\%) CL for 1 DOF, 
and the numbers in parentheses are the ones at 
3$\sigma$ (99.73\%) CL for 1 DOF. 
In the left, middle, and right columns, the input value of $\theta_{13}$ 
are taken as 
$\sin^2 2\theta_{13}=0.1$, 0.05, and 0.01, respectively. 
The column without number represents that no limit is obtained. 
}
\vglue 0.5cm
\begin{tabular}{c|ccc}
%\hline
%\multicolumn{3}{c}{54 km}\\
\hline
\ $\sigma_{usys}=1$\% \  &\ \ $\sin^2 2\theta_{13}=0.1$ \
             &\ \ $\sin^2 2\theta_{13}=0.05$ \
             &\ \ $\sin^2 2\theta_{13}=0.01$ \ \\
             \hline
\ Run type \  &\ \  \
             & $\delta(\Delta m^2) / \Delta m^2_{31}(0)$ (in \%)  at 1$\sigma$ (3$\sigma$) CL  
             &\ \  \\
\hline
Run I (5 locations)
                 &\ \ $^{+4.0}_{-3.6}$ $\left( ^{+12}_{-10} \right)$
                 &\ \ $^{+8.0}_{-7.2}$ $\left( ^{+27}_{-20} \right)$
                 &\ \  ---  \\
\hline
Run IIA (10 locations)
                 &\ \ 2.4 $\left( ^{+8.0}_{-7.2} \right)$
                 &\ \ 5.2 $\left( ^{+16}_{-14} \right)$
                 &\ \  $^{+26}_{-24}$ ( --- )  \\
\hline
Run IIB (10 locations)
                 &\ \ 1.2 (3.5) 
                 &\ \ 2.4 $\left( ^{+8.4}_{-7.2} \right)$
                 &\ \ $^{+15}_{-12}$   ( --- )  \\
\hline
Run III (20 locations)
                 &\ \ 0.8 (2.4) 
                 &\ \ 1.6 (5.2)
                 &\ \ $^{+9.2}_{-8.4}$  ( --- )  \\
\hline
\hline
\end{tabular}
\label{Tabmass_pessi}
%\vglue 0.5cm
\end{table}
%%%%%%%%%%%%%%%%%%%% Table III %%%%%%%%%%%%%%%%%%%%%%
%
%
%%%%%%%%%%%%%%%%%%%% Table IV %%%%%%%%%%%%%%%%%%%%
\begin{table}
\caption[aaa]{
The expected sensitivity to the $\sin^2 2\theta_{13}$ for the pessimistic  
systematic error of $\sigma_{usys} = 1$\% reachable by the 
Run I$-$III defined in the text. 
The uncertainties $\delta(\sin^2 2\theta_{13})$ 
are given at 1$\sigma$ (68.27\%) CL for 1 DOF, 
and the numbers in parentheses are the ones at 
3$\sigma$ (99.73\%) CL for 1 DOF. 
In the left, middle, and right columns, the input value of $\theta_{13}$ 
are taken as 
$\sin^2 2\theta_{13}=0.1$, 0.05, and 0.01, respectively. 
The column without number represents that no limit is obtained. 
}
\vglue 0.5cm
\begin{tabular}{c|ccc}
%\hline
%\multicolumn{3}{c}{54 km}\\
\hline
\ \ $\sigma_{usys}=1$\%  \  &\ \ $\sin^2 2\theta_{13}=0.1$ \
             &\ \ $\sin^2 2\theta_{13}=0.05$ \
             &\ \ $\sin^2 2\theta_{13}=0.01$ \\
             \hline
\ Run type \  &\ \  \
             &\ \ ~~$\sin^2 2\theta_{13}$ at 1$\sigma$ (3$\sigma$) CL~~ \
             &\ \  \ \\
\hline
Run I (5 locations)
                 &\ \ 0.1 $\pm 0.011$ (0.034) 
                 &\ \ 0.05 $\pm 0.012$ (0.037)
                 &\ \ 0.01 $^{+0.013}_{---}$ $\left( ^{+0.037}_{---} \right)$  \\
\hline
Run IIA (10 locations)
                 &\ \ 0.1 $\pm 0.009$ (0.026) 
                 &\ \ 0.05 $\pm 0.009$ $\left( ^{+0.027}_{-0.022} \right)$
                 &\ \ 0.01 $^{+0.009}_{---}$ $\left( ^{+0.028}_{---} \right)$  \\
\hline
Run IIB (10 locations)
                 &\ \ 0.1 $\pm 0.007$ (0.023)  
                 &\ \ 0.05 $\pm 0.008$ $\left( ^{+0.021}_{-0.024} \right)$
                 &\ \ 0.01 $\pm 0.008$ $\left( ^{+0.024}_{---} \right)$  \\
\hline
Run III (20 locations)
                 &\ \ 0.1 $\pm 0.006$ (0.017) 
                 &\ \ 0.05 $\pm 0.006$ (0.018)
                 &\ \ 0.01 $\pm 0.006$ $\left( ^{+0.019}_{---} \right)$  \\
\hline
\hline
\end{tabular}
\label{Tabangle_pessi}
\vglue 0.5cm
\end{table}

%%%%%%%%%%%%%%%%%%% Table IV %%%%%%%%%%%%%%%%%%

For disappearance measurement of 
$P(\bar{\nu}_{e} \rightarrow \bar{\nu}_{e} )$, 
$\sin^2 2\theta_{13}=0.01$ is a too small value 
for a pessimistic systematic error of $\sigma_{usys}=$ 1\% 
to retain the sensitivity to $\Delta m^2_{31}$. 
Therefore, reduction of the uncorrelated systematic error is 
the mandatory requirement  in this method for accurate measurement of 
$\Delta m^2_{31}$ at small $\theta_{13}$.

\subsection{Case without direct detection of $^3$H} 
\label{indirect}

%%%%%%%%%%%%%%%%%%%% Table V %%%%%%%%%%%%%%%%%%%%
\begin{table}
\caption[aaa]{
The expected fractional uncertainty 
$\delta(\Delta m^2) / \Delta m^2_{31}(0)$ in \%, and 
$\delta( \sin^2 2\theta_{13} )$ 
for the pessimistic systematic error of 
$\sigma_{usys} = 1$\%, 2\%, and 3\% 
reachable by Run 0 with 3 detectors as defined in the text. 
The uncertainties are given at 1$\sigma$ (68.27\%) CL for 1 DOF, 
and the numbers in parentheses are the ones at 
3$\sigma$ (99.73\%) CL for 1 DOF. 
In the left, middle, and right columns, the input value of $\theta_{13}$ 
are taken as 
$\sin^2 2\theta_{13}=0.1$, 0.05, and 0.01, respectively. 
The column without number represents that no limit is obtained. 
}
\vglue 0.5cm
\begin{tabular}{c|ccc}
%\hline
%\multicolumn{3}{c}{54 km}\\
\hline
\ \ Run 0  \  &\ \ $\sin^2 2\theta_{13}=0.1$ \
             &\ \ $\sin^2 2\theta_{13}=0.05$ \
             &\ \ $\sin^2 2\theta_{13}=0.01$ \\
\hline
\ $\sigma_{usys}$ \  &\ \  \
             & $\delta(\Delta m^2) / \Delta m^2_{31}(0)$ (in \%)  at 1$\sigma$ (3$\sigma$) CL \
             &\ \  \\
\hline
1 \% 
                 &\ \ $\left( ^{+8.6}_{-6.8} \right)$ ($\left( ^{+15}_{-13} \right)$) 
                 &\ \ $\left( ^{+12}_{-10} \right)$  (---)
                 &\ \ ---  (---)  \\
\hline
2 \% 
                 &\ \ $\left( ^{+12}_{-10} \right)$ (---) 
                 &\ \ $\left( ^{+18}_{-16} \right)$  (---)
                 &\ \ ---  (---)  \\
\hline
3 \% 
                 &\ \ $\left( ^{+15}_{-13} \right)$ (---) 
                 &\ \ ---  (---)
                 &\ \ ---  (---)  \\
\hline
\hline
\ $\sigma_{usys}$ \  &\ \  \
             &\ \ ~~$\sin^2 2\theta_{13}$ at 1$\sigma$ (3$\sigma$) CL~~ \
             &\ \  \ \\
\hline
1 \%
                 &\ \ 0.1 $\pm 0.012$ (0.035) 
                 &\ \ 0.05 $\pm 0.012$ (0.037)
                 &\ \ 0.01 $^{+0.013}_{---}$ $\left( ^{+0.038}_{---} \right)$  \\
\hline
2 \%
                 &\ \ 0.1 $\pm 0.022$ $\left( ^{+0.06}_{-0.07} \right)$ 
                 &\ \ 0.05 $\pm 0.024$ $\left( ^{+0.066}_{---} \right)$
                 &\ \ 0.01 $^{+0.024}_{---}$ $\left( ^{+0.069}_{---} \right)$  \\
\hline
3 \%
                 &\ \ 0.1 $\pm 0.033$ $\left( ^{+0.089}_{---} \right)$  
                 &\ \ 0.05 $\pm 0.035$ $\left( ^{+0.094}_{---} \right)$
                 &\ \ 0.01  $^{+0.035}_{---}$ $\left( ^{+0.098}_{---} \right)$  \\
\hline
\hline
\end{tabular}
\label{Tab-indirect}
\vglue 0.5cm
\end{table}

%%%%%%%%%%%%%%%%%%% Table V %%%%%%%%%%%%%%%%%%

Suppose that the direct detection of $^3$H in the target is not possible. 
Then, we may have to take the option of multiple detectors with the 
same structure, giving up the idea of movable detector. 
In this case,  most probably, we have to accept a pessimistic value 
of the uncorrelated systematic error of 1-3\%. 
It will cause two important changes in designing the experiment. 
(1) Number of events that can be accumulated in a 
reasonable time scale would be smaller by a factor of $\sim$10 
than the case of direct detection. 
(2) Number of detectors that can be prepared by keeping 
their identity to suppress the uncorrelated systematic errors 
may be limited. 
Therefore, 3 detector setting, for example, 
(in addition to a near detector which monitors the flux) 
would be more practical.

To understand performance of such reduced setting with larger errors, 
we have carried out the similar $\chi^2$ analysis as done in the 
previous subsections. 
We take 3 detector setting with tentatively determined baselines 
$L=\frac{1}{9} L_{\text{OM}}, L_{\text{OM}}$, and $3 L_{\text{OM}}$,  
and assume $10^{5}$ events in each detectors. 
We call the setting as Run 0. 
The three cases of the uncorrelated systematic errors, 1\%, 2\%, and 3\%, 
are examined. 
In Table~\ref{Tab-indirect}, presented are the expected fractional 
uncertainty $\delta(\Delta m^2) / \Delta m^2_{31}(0)$ and 
$\sin^2 2\theta_{13}$ for Run 0. 
With 1\% of the uncorrelated systematic errors, while a sensitivity 
comparable to Run I is reached for $\sin^2 2\theta_{13}$, 
uncertainty of $\Delta m^2_{31}$ is larger by a factor of $\simeq$2 
compared to Run I.
(Note that the baseline settings are not quite optimized in Run I.) 
For the cases of uncorrelated systematic errors of 2\% and 3\% the 
uncertainties of $\sin^2 2\theta_{13}$ get worse by a factor of 
$\simeq$2 and $\simeq$3, respectively. 
The behavior of sensitivities to $\Delta m^2_{31}$ are more 
complicated and no numbers are obtained for uncertainties at 
3$\sigma$ level for most cases. 
We note that loss of the sensitivities mainly comes from fewer number 
of detectors with larger systematic errors, but not from an order of 
magnitude smaller number of events.

The results obtained above indicate that the direct counting, 
either real time counting or an efficient extraction, of the 
produced $^3$H is mandatory to make the type of experiments 
under discussion useful.

\section{Analytic Estimation of the Sensitivities}\label{analytic}

We complement our numerical analysis of the sensitivities 
in the previous section by presenting analytical treatment 
of the uncertainties in the $\Delta m^2_{31}$ and 
$\theta_{13}$ determination. 
In particular, we derive analytic formulas for the sensitivities of 
$\Delta m^2_{31}$ and $\sin^2 2\theta_{13}$ under the approximation 
of small uncorrelated systematic error compared to correlated one, 
$\sigma^2_{u} \ll \sigma^2_{c}$. 
In Sec.~\ref{method}, we have argued, 
assuming feasible direct counting of $^3$H atoms, 
that the hierarchy of errors is very likely to hold.

We restrict ourselves, 
in consistent with the numerical analysis done in the previous section, 
to the case that an equal number of events are taken in each baseline, 
$N^{obs}_{i} = N^{obs}$, which may be translated into $N^{exp}_{i} = N^{exp}$. 
We also assume, for simplicity, the case of equal uncorrelated 
systematic error in each detector location, 
$\sigma^2_{u i} \equiv \sigma^2_{usys,i} + \frac{1}{N^{exp}_{i}} = \sigma^2_{u}$. 
Under these assumptions, $V^{-1}$ has a simple form 
\begin{eqnarray}
V^{-1} = 
\frac{1}{\sigma^2_{u}} 
\left[  I - 
\frac{1}{n}
H_{n \times n}
\right]  
= \frac{1}{\sigma^2_{u}} 
\left(\begin{array}{ccccc}
1 - \frac{1}{n} &-\frac{1}{n}& -\frac{1}{n} &\cdots & -\frac{1}{n} \\
-\frac{1}{n} &1 - \frac{1}{n}  & -\frac{1}{n} &\cdots & -\frac{1}{n} \\
-\frac{1}{n} &-\frac{1}{n} & 1 - \frac{1}{n}  &\cdots & -\frac{1}{n} \\
\vdots&\vdots&\vdots&\vdots&\vdots\\
-\frac{1}{n} &-\frac{1}{n} & -\frac{1}{n} &\cdots &1 - \frac{1}{n}  \\
\end{array}
\right), 
\label{Vinv_matrix}
\end{eqnarray}
where $H_{n \times n}$ is an $n \times n$ matrix whose elements 
are all unity, $H_{i, j} = 1$ for any $i$ and $j$. 
It indicates again the independence of the $\chi^2$ on the correlated error 
$\sigma^2_{c}$ and the ``scaling behavior'' with respect to 
the uncorrelated systematic error. 
Then, the $\chi^2$ simplifies:
\begin{eqnarray}
\chi^2 = 
\frac{1}{n \sigma^2_{u}}
\sum_{i, j =1}^{n} \left( 
\frac{N^{obs}_{i}}{N^{exp}_{i}} - 
\frac{N^{obs}_{j}}{N^{exp}_{j}}
\right)^2. 
\label{chi2_nxn}
\end{eqnarray}

\subsection{Optimal baselines and sensitivities for two detector locations}\label{optimal_eqn}

Let us start by examining sensitivities for the case of two detector locations. 
Because of a simple setting with monochromatic $\bar{\nu_{e}}$ beam 
we can give explicit expression of $\chi^2$ in terms of small 
deviation of the parameters from the true (nature's) values. 
For this purpose, we note that the number of events is given by 
\begin{eqnarray}
N^{exp} (N^{obs}) = f_{\bar{\nu_{e}}}  \sigma_{res} N_{T} T 
P_{ee} (\theta_{13}, \Delta m^2_{31}, L), 
\label{number}
\end{eqnarray}
where 
$f_{\bar{\nu_{e}}}$ denotes the neutrino flux, 
$\sigma_{res}$ the absorption cross section, 
$N_{T}$ the number of target nuclei, 
$T$ the running time, 
and 
$P_{ee}$ is a short-hand notation for 
$P(\bar{\nu_{e}} \rightarrow \bar{\nu_{e}})$.
In the setting in this section, 
$T$ is adjusted such that an equal number of events is collected 
at each location of the detector.  
We recall that $N^{obs}$ denotes the event number computed 
with the true value of the parameters, 
$\Delta m^2_{31}=\Delta m^2(0)$ and 
$\sin^2 2\theta_{13}=\sin^2 2\theta_{13}(0)$, 
whereas 
$N^{exp}$ denotes the event number computed 
with possible small deviations $\delta(\Delta m^2)$ and 
$\delta(\sin^2 2\theta_{13})$ from the true values of the parameters. 
Then, $i$-th component of $\vec{x}$ vector in (\ref{vector}) is 
given to first order in the deviation as 
\begin{eqnarray}
\frac{N^{obs}_{i}}{N^{exp}_{i}} - 1 &=&  
\frac{ 1 }
{ P_{ee}^{(0)}(L_{i}) }
\left[ 
\sin^2 {\left(
\frac{\Delta m^2_{31}(0) L_{i}}{4 E}
\right)}
\delta(\sin^2 2\theta_{13}) 
\right.
\nonumber \\
\hspace*{0mm} {}&+&
\left.
%+ 
\frac{1}{2}
\sin^2 2\theta_{13}(0) 
\sin {\left(
\frac{\Delta m^2_{31}(0) L_{i}}{2 E}
\right)} \times 
\left(
\frac{\delta( \Delta m^2) L_{i}}{2 E}
\right)
\right], 
\label{x-comp}
\end{eqnarray}
where 
$P_{ee}^{(0)}(L_{i}) \equiv 
P_{ee} [\theta_{13}(0), \Delta m^2_{31}(0), L_{i}]$ for which 
we have used the two-flavor expression (\ref{Pee}).

For simplicity we restrict our discussion in this section to the analysis 
with single degree of freedom. It is a natural setting for estimating 
ultimate sensitivities; 
When we discuss sensitivity of $\Delta m^2_{31}$ we optimize 
$\chi^2$ in terms of $\sin^2 2\theta_{13}$, and vice versa. 
Or, one can think of the situation that, 
 in determination of $\Delta m^2_{31}$, 
$\sin^2 2\theta_{13}$ is accurately determined by other ways, 
e.g., long-baseline accelerator experiments. 
Under the approximation $\sigma^2_{u} \ll \sigma^2_{c}$ and 
using $V^{-1}$ in (\ref{Vinv_matrix}), $\chi^2$ is given 
for small deviations of $\theta_{13}$ and $\Delta m^2_{31}$ as follows:
\begin{eqnarray}
\chi^2_{\theta} = 
\frac{[\delta(\sin^2 2\theta_{13})]^2}{2 \sigma^2_{u}} 
\left[
\frac{ \sin^2 \left( \frac{\Delta m^2_{31}(0) L_{1}}{4 E} \right) }
{ P_{ee}^{(0)}(L_{1}) } - 
\frac{ \sin^2 \left( \frac{\Delta m^2_{31}(0) L_{2}}{4 E} \right) }
{ P_{ee}^{(0)}(L_{2}) }
\right]^2, 
\label{chi2_theta}
\end{eqnarray}
\begin{eqnarray}
\chi^2_{\Delta m^2} &=& 
\frac{[\sin^2 2\theta_{13}(0)]^2}{8 \sigma^2_{u}} 
\left( \frac{\delta(\Delta m^2)}{\Delta m^2_{31}(0)} \right)^2 
\nonumber \\
&\times&
\left[
\frac{ \left( \frac{\Delta m^2_{31}(0) L_{1}}{2 E} \right)  
\sin \left( \frac{\Delta m^2_{31}(0) L_{1}}{2 E} \right) }
{ P_{ee}^{(0)}(L_{1}) }  - 
\frac{ \left( \frac{\Delta m^2_{31}(0) L_{2}}{2 E} \right)  
\sin \left( \frac{\Delta m^2_{31}(0) L_{2}}{2 E} \right) }
{ P_{ee}^{(0)}(L_{2}) } 
\right]^2. 
\label{chi2_dm2}
\end{eqnarray}

Now, we can address the problem of optimal baseline and 
estimate the sensitivities of $\sin^2 2\theta_{13}$ 
and $\Delta m^2_{31}$ under the approximations stated above. 
Since $\sin^2 2\theta_{13} \lsim 0.1$ \cite{CHOOZ} 
it may be a reasonable approximation to set 
$P_{ee}^{(0)}(L_{i}) =1$ 
in the denominator, as we do in the rest of the section.

\subsubsection{Optimal setting and sensitivity to $\sin^2 2\theta_{13} $}

To maximize (\ref{chi2_theta}) one should take $L_{1}$ as short as possible, 
and $L_{2}$ at the oscillation maximum, the well known feature in the 
reactor $\theta_{13}$ experiments. 
Thus, we take $L_{1}=0$ and $L_{2} = L_{\text{OM}} $ which makes the square 
parenthesis in (\ref{chi2_theta}) unity. 
Then, one can obtain the sensitivity at $N_{CL} \sigma$ CL for 
2 detector locations as 
\begin{eqnarray}
\delta(\sin^2 2\theta_{13}) =  
2 N_{CL} \frac{\sigma_{u}}{\sqrt{2}} =
2 N_{CL} 
\sqrt{ 
\frac{\sigma_{usys}^2 + \frac{1}{N} }{2}  }. 
\label{sens_theta}
\end{eqnarray}

For $\sigma_{u} = 0.2$\%, 
$\delta(\sin^2 2\theta_{13}) = 2.8 \times 10^{-3}$ ($8.5 \times 10^{-3}$) 
at 1$\sigma$ (3$\sigma$) CL. 
For $\sigma_{u} = 1$\%, 
$\delta(\sin^2 2\theta_{13}) = 0.014$ (0.043) 
at 1$\sigma$ (3$\sigma$) CL, which is not so far from the 
sensitivities quoted in the literatures of the reactor $\theta_{13}$
experiments.

\subsubsection{Optimal setting and sensitivity to $\Delta m^2_{31}$}

The optimal baseline setting is quite different for $\Delta m^2_{31}$. 
We first recall a property of the function $x \sin x$;  
It has the first maximum $x \sin x = 1.82$ at $x = 2.02$ ($L= 0.64 L_{\text{OM}} $),  
has the first minimum $x \sin x = - 4.81$ at $x = 4.91$ ($L= 1.56 L_{\text{OM}} $), 
and then, 
the second maximum $x \sin x = 7.92$ at $x = 7.98$ ($L= 2.54 L_{\text{OM}} $), 
and so on. 
For simplicity, we restrict ourselves to $x \leq 3\pi$, which means 
$L \leq 3 L_{\text{OM}}$ so that the running time does not blow up. 
Then,  the optimal setting is 
$L_{1}= 0.64 L_{\text{OM}} $ and $L_{2} = 1.56 L_{\text{OM}} $ 
if we restrict to $L \leq 2 L_{\text{OM}}$, and 
$L_{1} = 1.56 L_{\text{OM}} $ and $L_{2}= 2.54 L_{\text{OM}} $ 
if we allow baseline until $L \leq 3 L_{\text{OM}}$. 
Despite the factor of $\simeq$4 different baseline lengths, we still 
assume the equal numbers of events at 
$L=0.64 L_{\text{OM}} $ and $L=2.54 L_{\text{OM}} $, 
which implies $\simeq$16 times longer exposure time at the latter distance.

The $\chi^2$ is given approximately by 
\begin{eqnarray}
\chi^2_{\Delta m^2} = 
%\left( \frac{n}{2} \right) 
\frac{c}{\sigma^2_{u}}
\sin^4 2\theta_{13}(0)
\left( \frac{\delta(\Delta m^2_{31})}{\Delta m^2_{31}(0)} \right)^2. 
\label{chi2_dm22}
\end{eqnarray}
The coefficient $c$ is 
5.5 for $L \leq 2 L_{\text{OM}}$ and 
20.3 for $L \leq 3 L_{\text{OM}}$. 
Then, we obtain the sensitivity at $N_{CL} \sigma$ CL for 2 detector locations as 
\begin{eqnarray}
\frac{\delta(\Delta m^2_{31})}{\Delta m^2_{31}(0)} = 
\frac{ 1  } { \sin^2 2\theta_{13}(0) } 
N_{CL} 
\frac{\sigma_{u}}{\sqrt{c}}, 
\label{sens_dm2}
\end{eqnarray}
Hence, the sensitivity to $\Delta m^2_{31}$ depends very sensitively 
on $\sin^2 2\theta_{13}$.

With $\sigma_{u} = 0.2$\%, 
$\frac{\delta(\Delta m^2_{31})}{\Delta m^2_{31}(0)} = 8.5 \times 10^{-3}$ 
at 1$\sigma$ CL for $\sin^2 2\theta_{13}=0.1$ if we restrict to 
$L \leq 2 L_{\text{OM}}$. 
If we allow $L \leq 3 L_{\text{OM}}$, the sensitivity becomes better,  
$\frac{\delta(\Delta m^2_{31})}{\Delta m^2_{31}(0)} = 4.4 \times 10^{-3}$ 
at the same CL.
If the systematic error is worse, $\sigma_{u} = 1$\%, 
the sensitivity at $\sin^2 2\theta_{13}=0.1$ becomes to 
$\frac{\delta(\Delta m^2_{31})}{\Delta m^2_{31}(0)} = 4.3 \times 10^{-2}$ and  
$\frac{\delta(\Delta m^2_{31})}{\Delta m^2_{31}(0)} = 2.2 \times 10^{-2}$ at 1$\sigma$ CL 
for $L \leq 2 L_{\text{OM}}$ and $L \leq 3 L_{\text{OM}}$ cases, 
respectively.

\subsection{The problem of $n$ detector locations reduces to 2 location case}

We first show that the problem of optimal setting of $n$ detector 
locations reduces to the case of 2 locations under the assumption 
of equal number of events in each location. 
To indicate the essential point let us first consider a simplified $\chi^2$ 
of the form 
$\chi^2 = \sum_{i, j = 1}^{n} (x_{i} - x_{j})^2$ and $0 \leq x_{i} \leq 1$, 
which is the essential part of $\chi^2$ for $\sin^2 2\theta_{13}$, the equation 
(\ref{chi2_theta}). 
In the case of 2 locations the configuration which maximizes 
the $\chi^2(n=2)$ is $x_{1} = 0$ and $x_{2} = 1$ and 
$\chi^2_{max}(n=2)=1$. 
It is not difficult to observe that in the case of $n$-locations 
the configuration which maximizes $\chi^2(n)$ is 

\begin{itemize}

\item

Even n:  n=2M; x=0 appears M times, and x = 1 appears M times, 
%\\
$\chi^2 (n)_{max} = M^2$.

\item

Odd n:  n=2M+1; x=0 appears M+1 times, and x = 1 appears M times, or vice versa, \\
$\chi^2 (n)_{max} = M(M+1)$.

\end{itemize}
Thus, the problem of optimal setting with equal number of events 
at each detector location is reduced to the 2 location case.

For $\chi^2$ for $\Delta m^2_{31}$ in (\ref{chi2_dm2}) the situation 
is slightly different because the function $|x \sin x|$ increases 
without limit as $x$ becomes large. 
Therefore, mathematically speaking, one can obtain better and 
better accuracies as one goes to longer and longer distances 
in our setting of equal number of events at any detector location.\footnote{
%%%%%%%%%%%%%%%% \footnote %%%%%%%%%%%%%%%%
This explains at least partly the reason why the sensitivities to 
$\Delta m^2_{31}$ and $\sin^2 2\theta_{13}$ differ in dependence on 
distance from the source to a detector, as indicated in Fig.~3 in \cite{SADO}. 
}
But, since we want to remain to a reasonable running time, we 
have restricted our discussions to baselines limited by 
$L \leq 2 L_{\text{OM}}$ or $L \leq 3 L_{\text{OM}}$ 
in the 2 location case, the restriction which is kept throughout this section. 
Then, one can show that the same result follows for 
$\chi^2$ for $\Delta m^2_{31}$, (\ref{chi2_dm2}). 
Namely, in the case of $n$ locations, the highest sensitivity is 
achieved at the same baselines $L_{1}$ and $L_{2}$ of the 2 location case; 
$L_{1}$ in $\left[ \frac{n}{2} \right]$ 
($\left[ \frac{n}{2} \right] + 1$ for odd $n$) times, and 
$L_{2}$ in $\left[ \frac{n}{2} \right]$ times, where $[~]$ implies Gauss' symbol.

The maximal value of $\chi^2$ is, therefore, given by 
\begin{eqnarray}
\chi^2 (n)_{max} = 
\left( \frac{n}{2} \right) \chi^2 (n=2)_{max}; (\text{even}~ n), 
\nonumber \\
\chi^2 (n)_{max} = 
\left( \frac{n^2-1}{2 n} \right) \chi^2 (n=2)_{max}; (\text{odd}~ n), 
\label{chi2-n}
\end{eqnarray}
It can be translated into the uncertainties 
at $N_{CL}$ CL as  
\begin{eqnarray}
\delta(\sin^2 2\theta_{13}) (n) &=& 
\sqrt{ \frac{2}{n} }
\delta(\sin^2 2\theta_{13}) (n=2), 
\nonumber \\
\left( \frac{\delta(\Delta m^2_{31})}{\Delta m^2_{31}(0)} \right) (n)  &=& 
\sqrt{ \frac{2}{n} }
\left( \frac{\delta(\Delta m^2_{31})}{\Delta m^2_{31}(0)} \right) (n=2), 
\label{sensitivities}
\end{eqnarray}
for even $n$. 
For odd $n$, $n$ in (\ref{sensitivities}) must be replaced by $\frac{n^2 - 1}{n}$. 
$\delta(\sin^2 2\theta_{13}) (n=2)$ and 
$\left( \frac{\delta(\Delta m^2_{31})}{\Delta m^2_{31}(0)} \right) (n=2)$
are given respectively by (\ref{sens_theta}) and (\ref{sens_dm2}). 
Therefore, the sensitivity gradually improves as number of runs 
becomes larger.\footnote{
%%%%%%%%%%%%% footnote %%%%%%%%%%%%%%%
It is the well known feature in the multiple detector setting in the 
reactor $\theta_{13}$ experiments in which one obtains better 
sensitivity as in (\ref{sens_theta})  
if the the two identical detectors, the near and the far, 
are each divided into small detectors in the same way, 
if the uncorrelated systematic error $\sigma_{usys}$ is made to be 
equal with that of the original large detector and if the statistical 
errors are negligible even for divided detectors. 
This point was emphasized by Yasuda \cite{yasuda}.  
}
%%%%%%%%%%%%% footnote %%%%%%%%%%%%%%%

At the end of this subsection, we want to note the followings:
The reason why we did not take these sets of the optimal distances for 
 $\theta_{13}$ and $\Delta m^2_{31}$ obtained in this  
subsection  in the numerical analyses in 
Sec.~\ref{numerical} is that the sensitivity to 
$\Delta m^2_{31}$ is lost if we tune the setting optimal for 
$\sin^2 2\theta_{13}$, and vice versa. 
The reason for this is easy to understand;
At baselines $L_{1}$ and $L_{2}$ which maximizes 
$\chi^2_{\theta}$ ($\chi^2_{\Delta m^2}$), 
$\chi^2_{\Delta m^2}$ ($\chi^2_{\theta}$) vanishes 
(approximately vanishes), because 
$\sin x_1 = \sin x_2 = 0$ at $x_1 =0$ and $x_2 = \pi$ 
($\sin^2 \frac{x_1}{2} \simeq \sin^2 \frac{x_2}{2} \simeq \frac{1}{2}$ at 
$x_1 \simeq \frac{\pi}{2}$ and $x_2 \simeq \frac{3 \pi}{2}$). 
Nonetheless, we will see, in the following subsections, that the 
sensitivities analytically estimated with optimal baseline distances 
and the numerically calculated ones with baselines taken by 
``common sense'' agree reasonably well with each other.

\subsection{Analytic estimation of the sensitivities; $\sin^2 2\theta_{13}$ }

Let us examine the case of $\sigma_{usys}=0.2$\% and 
the number of events $N=10^{6}$ which was considered in our numerical 
analysis in Sec.~\ref{numerical}. Then, $\sigma_{u}=0.22$\%. 
In the case of 5 locations, $n=5$, 
$\delta(\sin^2 2\theta_{13}) = 2.0 \times 10^{-3}$ 
($6.1 \times 10^{-3}$)
at 1$\sigma$ (3$\sigma$) CL.
Similarly for 10 and 20 locations 
$\delta(\sin^2 2\theta_{13}) = 1.4 \times 10^{-3}$ 
($4.2 \times 10^{-3}$) and 
$0.99 \times 10^{-3}$ ($3.0 \times 10^{-3}$), 
respectively, 
at 1$\sigma$ (3$\sigma$) CL.
They compare well with the numbers in Table~\ref{Tabangle} 
though the latter are obtained with not-so-tuned baseline settings.
Notice that 
$\delta(\sin^2 2\theta_{13})$ is independent of $\theta_{13}$ 
under the present approximation of small deviation from the best fit.

With $\sigma_{usys}=1$\% the corresponding sensitivities are 
$\delta(\sin^2 2\theta_{13}) = 9.2 \times 10^{-3}$ 
($2.8 \times 10^{-2}$), 
$6.4 \times 10^{-3}$ ($1.9 \times 10^{-2}$), and  
$4.5 \times 10^{-3}$ ($1.3 \times 10^{-2}$) 
for 5, 10, and 20 locations, respectively, 
at 1$\sigma$ (3$\sigma$) CL.
They are again roughly consistent with the ones in Table~\ref{Tabangle_pessi}.

\subsection{Analytic estimation of the sensitivities; $\Delta m^2_{31}$ }

We examine the cases of restriction $L \leq 2 L_{\text{OM}}$ 
and $L \leq 3 L_{\text{OM}}$. 
In the first case, the maximum of the function $x \sin x$ 
is at $x = 2.02$ ($L= 0.64 L_{\text{OM}} $) where $x \sin x = 1.82$, and 
the minimum at $x = 4.91$ ($L= 1.56 L_{\text{OM}} $) 
where $x \sin x = - 4.81$. 
In the case of milder restriction $L \leq 3 L_{\text{OM}}$, 
the maximum of the function $x \sin x$ 
is at $x = 7.98$ ($L= 2.54 L_{\text{OM}} $) where $x \sin x = 7.92$ , and 
the minimum at $x = 4.91$ ($L= 1.56 L_{\text{OM}} $) as above. 
%where $x \sin x = - 4.81$. 

In Table~\ref{Tabmass_analytic}, we give the fractional uncertainties 
of $\Delta m^2_{31}$, 
$\delta(\Delta m^2) / \Delta m^2_{31}(0)$ in \% at 1$\sigma$ CL for  
the optimistic and the pessimistic systematic errors of 
$\sigma_{usys}=0.2$\% and 1\% (the latter in parenthesis), respectively, 
obtained by using the equations 
(\ref{sens_dm2}) and (\ref{sensitivities}). 
We do not show errors at 3$\sigma$ CL because it is obtained 
simply by multiplying 3. 
Over-all, the analytically estimated uncertainties are in reasonable 
agreement with those obtained by the numerical analysis in 
Sec.~\ref{numerical}. 
Notice that one has to compare the sensitivities of 
Run I and IIA with the case of severer restriction 
$L \leq 2 L_{\text{OM}}$, and the ones of Run IIB and III with
the case of milder restriction $L \leq 3 L_{\text{OM}}$, because 
distances beyond $2 L_{\text{OM}}$  are involved in the latter runs. 
The fact that our analytical estimates of the 
errors are smaller than the numerical ones by $\sim$30\% or so,  
apart from approximations involved, is consistent 
with that the latter are based on non-optimal baseline distances. 
It also implies that the baseline setting chosen by the  ``common sense'' 
used in the numerical analysis in Sec.~\ref{numerical} is not so far 
from the optimal one, indicating that the sensitivities are rather stable 
against changes of baseline setting.

%%%%%%%%%%%%%%%%%%%% Table VI %%%%%%%%%%%%%%%%%%%%%
\begin{table}
\caption[aaa]{
The analytically estimated fractional uncertainties of $\Delta m^2_{31}$, 
$\delta(\Delta m^2) / \Delta m^2_{31}(0)$ in \% are given for  
the optimistic systematic error of $\sigma_{usys} = 0.2$\% 
and for the pessimistic one (in parentheses) of $\sigma_{usys} = 1$\%. 
The uncertainties are given at 1$\sigma$ (68.27\%) CL for 1 DOF 
for the cases of 5, 10, and 20 detector locations. 
The upper three lows are for the case of restricted baselines, 
$L \leq 2 L_{\text{OM}}$, 
whereas the lower three lows are for the cases of somewhat 
relaxed baseline setting, $L \leq 3 L_{\text{OM}}$. 
In the left, middle, and right columns, the input value of $\theta_{13}$ 
are taken as 
$\sin^2 2\theta_{13}=0.1$, 0.05, and 0.01, respectively.   
}
\vglue 0.5cm
\begin{tabular}{c|ccc}
%\hline
%\multicolumn{3}{c}{54 km}\\
\hline
\ $L \leq 2 L_{\text{OM}}$ \  &\ \  \
             &\ \ $\sigma_{usys}=0.2$\% ($\sigma_{usys}=1$\%)  \
             &\ \  \ \\
\hline
\ number of locations \  &\ \  \
             & $\delta(\Delta m^2) / \Delta m^2_{31}(0)$ (in \%)  at 1$\sigma$CL \ 
             &\ \  \\
\hline
\   \  &\ \ $\sin^2 2\theta_{13}=0.1$ \
             &\ \ $\sin^2 2\theta_{13}=0.05$ \
             &\ \ $\sin^2 2\theta_{13}=0.01$ \ \\
             \hline
5 locations 
                 &\ \ 0.61 (2.8) 
                 &\ \ 1.2  (5.5)
                 &\ \ 6.1  (28)  \\
\hline
10 locations 
                 &\ \ 0.42 (1.9) 
                 &\ \ 0.85  (3.8)
                 &\ \ 4.2  (19) \\
\hline
20 locations 
                 &\ \ 0.30  (1.3) 
                 &\ \ 0.6  (2.7)
                 &\ \ 3.0  (13) \\
\hline
\hline
\  $L \leq 3 L_{\text{OM}}$ \  &\ \  \
             & $\delta(\Delta m^2) / \Delta m^2_{31}(0)$  (in \%)  at 1$\sigma$CL   \
             &\ \  \ \\
 \hline
 5 locations 
                 &\ \ 0.32 (1.5) 
                 &\ \ 0.62 (2.9)
                 &\ \ 3.2 (15) \\
\hline
10 locations 
                 &\ \ 0.22 (0.99) 
                 &\ \ 0.44 (2.0)
                 &\ \ 2.2 (9.9) \\
\hline
20 locations 
                 &\ \ 0.15 (0.68) 
                 &\ \ 0.31 (1.4)
                 &\ \ 1.5 (6.8) \\
\hline
\hline
\end{tabular}
\label{Tabmass_analytic}
%\vglue 0.5cm
\end{table}
%%%%%%%%%%%%%%%%%%%% Table VI %%%%%%%%%%%%%%%%%%%%%

Based on the numerical and the analytical estimate of the uncertainties 
of $\Delta m^2_{31}$ determination, we conclude that sensitivities of 
less than 1\% at 90\% CL required for resolution 
of mass hierarchy proposed in \cite{NPZ,GJK} is in reach 
if the uncorrelated systematic error of $\sigma_{usys}=0.2$\% is 
realized, and if $\theta_{13}$ is relatively large, 
$\sin^2 2\theta_{13} \gsim 0.05$ in Run IIB.  

%The present method may {\it not} work for so small $\theta_{13}$ that conventional superbeam experiments would not reach. 

We have confined ourselves to the problem of optimal setting of 
distances under the constraint of equal number of events at each location. 
To minimize running time for a given sensitivity, 
we have to address the problem of optimal detector locations and 
exposure times for a given total running time. It is left for a future study.

%\newpage

\section{Concluding remarks}\label{conclusion}

In this paper, we have explored the potential of high sensitivity 
measurement of $\Delta m^2_{31}$ and $\theta_{13}$ which is 
enabled by using the resonant absorption of monochromatic 
$\bar{\nu}_{e}$ beam enhanced by the M\"ossbauer effect. 
With baseline distances of $\sim$10 m, the movable detector setting 
is certainly possible. 
Assuming that the direct detection of produced $^3$H atom either 
by real time counting or extraction of $^3$H atoms works, 
we have argued that the uncorrelated systematic error can be 
as small as 0.1\%-0.3\%, if there equipped a near detector with the 
same structure with the far one. 
It will allow us to determine $\Delta m^2_{31}$ 
%to less than 1\% level at 1$\sigma$ CL for $\sin^2 2\theta_{13} \geq 0.03$ 
to the accuracies $\Delta m^2_{31}$ of 
$\simeq0.3~(\sin^2 2\theta_{13} / 0.1)^{-1}$ \% at 1$\sigma$ CL 
(Run IIB, $\sigma_{usys}=0.2$\%). 
The error of $\sin^2 2\theta_{13}$ is also small, 
$\sin^2 2\theta_{13}=1.8 \times 10^{-3}$ almost independently 
of $\theta_{13}$ with the same setting. 
The accuracy of the $\theta_{13}$ measurement even in Run I, 
if the systematic error of 0.2\% is reached, is comparable with that 
of the next generation accelerator $\nu_{e}$ appearance experiments 
\cite{JPARC,NOVA}. 
It may exceed the accelerator sensitivity if Run IIB is performed. 
If the systematic error is of 1\% level, the sensitivity to $\theta_{13}$ 
is similar to the first-stage reactor $\theta_{13}$ experiments 
even in Run IIB.

What is the scientific merit of the precision measurement of 
$\Delta m^2_{31}$ and $\theta_{13}$?
As we have already mentioned in Sec.~\ref{introduction} the precision 
measurement of $\Delta m^2_{31}$ and $\theta_{13}$ will have a 
great impact, at least, to the two of the unknowns in the lepton flavor 
mixing, the neutrino mass hierarchy and resolving the $\theta_{23}$ 
octant degeneracy. 
It would be very interesting to carry out quantitative analyses of 
such possibilities \cite{MNPZ}.
We emphasize that such physics capabilities can only 
be made possible by the direct counting of the produced $^3$H atoms 
which makes movable detector feasible. 
We hope that these exciting possibilities stimulate further development 
of the experimental technology toward the goal.

What is the additional capabilities of the resonant absorption of monochromatic 
$\bar{\nu}_{e}$ beam? 
With 18.6 keV of neutrino energy the solar oscillation maximum 
would be reached at 
$L_{\text{solarOM}} = 290  
\left(  \frac{\Delta m^2_{21}}{8 \times 10^{-5} \text{eV}^2}  \right)^{-1} 
\text{m}$. 
Then, it would be worthwhile to explore the possibility of precision 
measurement of $\theta_{12}$ and $\Delta m^2_{21}$, 
as was done for the reactor experiments \cite{SADO}. 
But, in the present case, neither geo-neutrinos nor 
$\bar{\nu_{e}}$ flux from nearby reactors (if any) contaminate 
the measurement. 
In particular, possible movable or multiple baseline set up should 
allow improvement of accuracy of $\Delta m^2_{21}$ determination.

Detection of CP violating effect due to $\delta$ in the 
$\bar{\nu}_{e}$ disappearance measurement requires to go down 
by a factor of $O(10^{-6})$ compared to CP conserving terms \cite{solarCP}, 
Unfortunately, it would not be in reach despite great potential sensitivities 
achievable by the resonant absorption reaction.

Finally, the setup of multiple baseline lengths of $\sim$10 m allows a 
precision test of the pure vacuum oscillation hypothesis by observing 
a sine curve slightly modified by the solar $\Delta m^2_{21}$ oscillation. 
It will constrain various possible sub-leading effects such as 
de-coherence or new neutrino interactions to a great precision.

%\newpage

\appendix

\section{General formula for $\chi^2$ for $n$-uncorrelated and 
$\ell$-correlated errors\label{appendix1}}

We consider a general setting with $n$-uncorrelated and $\ell$-correlated errors. 
$\chi^2$ is given in a form as
\begin{eqnarray}
\chi^2 = 
\vec{x}^{\text{T}} V^{-1} \vec{x} 
\label{chi21}
\end{eqnarray}
where
$\vec{x}^{T} = (x_{1}, x_{2}, ... ,x_{n})$ and 
$x_i = \frac{N^{obs}_{i}}{N^{exp}_{i}} -1$.
We also introduce a vector notation for parameters 
$\alpha_{p}$ ($p=1, .. \ell$) for correlated error as 
$\vec{\alpha}^{T} = (\alpha_{1}, \alpha_{2}, ... \alpha_{\ell})$ 
for so called the ``pull type'' $\chi^2$ \cite{pull,pull-original}. 
Then, $\chi^2$ can be written \cite{SYSHS} 
\begin{eqnarray}
\chi^2 &=& \min_{\vec{\alpha}}
\left[\left(\vec{x}-H\vec{\alpha}\right)^T D_{\text{u}}^{-1}
\left(\vec{x}-H\vec{\alpha}\right)
+\vec{\alpha}^T D_{\text{c}}^{-1}\vec{\alpha}
\right]
\nonumber\\
&=& 
\min_{\vec{\alpha}}\left[
\left(\vec{\alpha}-A^{-1}H^{T}D_{\text{u}}^{-1}\vec{x}\right)^T
A\left(\vec{\alpha}-A^{-1}H^{T}D_{\text{u}}^{-1}\vec{x}\right)
+\vec{x}^T \left(D_{\text{u}}^{-1}-D_{\text{u}}^{-1}
HA^{-1}H^TD_{\text{u}}^{-1}\right)\vec{x}
\right].
\nonumber\\
\label{pull1}
\end{eqnarray}
After minimization by $\alpha_{j}$, $V^{-1}$ can be written as 
\begin{eqnarray}
V^{-1}=D_{\text{u}}^{-1}-D_{\text{u}}^{-1}
HA^{-1}H^TD_{\text{u}}^{-1}.
\label{Vinv}
\end{eqnarray}
In the above equations,  
$D_{u}$ is an $n \times n$ matrix 
$D_{u} \equiv \mbox{\rm diag} 
\left(
\sigma_{u1}^2,\cdots,\sigma_{un}^2
\right)$, 
$D_{c}$ is an $\ell \times \ell$ matrix 
$D_{c} \equiv \mbox{\rm diag} 
\left(
\sigma_{c1}^2,\cdots,\sigma_{c \ell}^2
\right)$, 
and 
$H$ is an  $n \times \ell$ matrix whose elements are all unity, 
$H_{p, i} =1$ for any $p=1, ... \ell$ and $i = 1, ... n$. 
Finally, the $\ell \times \ell$ matrix $A$ is defined by 
\begin{eqnarray}
A\equiv D_{\text{c}}^{-1}+H^TD_{\text{u}}^{-1}H.
\label{A}
\end{eqnarray}
Note that the statistical errors are incorporated in the uncorrelated errors as 
$\sigma_{ui}^2 = \sigma_{sys; ui}^2 + \frac{1}{N^{exp}_{i}}$.
A simple ``path-integral'' proof is given by  \cite{SYSHS} that 
$V=D_{\text{u}}+HD_{\text{c}}H^T$.

Here, we are interested in obtaining the explicit form of $V^{-1}$ in (\ref{Vinv}). 
We first compute $A^{-1}$. We note that 
the matrix $A$ given in (\ref{A}) can be written explicitly as 
\begin{eqnarray}
A = 
\left(
\sum_{i=1}^{n} \frac{1}{\sigma^2_{u i}}
\right) T,
\label{Akl}
\end{eqnarray}
\begin{eqnarray}
T = \left(\begin{array}{ccccc}
1+ \epsilon_{1} &1& 1 &\cdots & 1 \\
1 &1+ \epsilon_{2} & 1 &\cdots & 1 \\
1 &1 & 1 + \epsilon_{3} &\cdots & 1 \\
\vdots&\vdots&\vdots&\vdots&\vdots\\
1 &1 & 1 &\cdots & 1+ \epsilon_{\ell}  \\
\end{array}
\right),
\label{T}
\end{eqnarray}
where $\epsilon_{i}$ in (\ref{T}) is given by 
\begin{eqnarray}
\epsilon_{p} = 
\frac{1}{\sigma^2_{c p}} 
\left(
\sum_{i=1}^{n} \frac{1}{\sigma^2_{u i}}
\right)^{-1}
\hspace{1cm}
(p=1, ... \ell).
\label{epsilon}
\end{eqnarray}
It is easy to show that $T^{-1}$ is given as
\begin{eqnarray}
(T^{-1})_{pp} &=&  
\frac{1}{\epsilon_{p}}
\frac{ \left[
1 + \sum_{r \neq p}^{\ell}  \frac{1}{\epsilon_{r} }
\right] }
{\left[
1 + \sum_{r=1}^{\ell}  \frac{1}{\epsilon_{r} }
\right] }  = 
\frac{1}{\epsilon_{p}} - 
\frac{ 1 }
{\epsilon_{p}^2 \left[
1 + \sum_{r=1}^{\ell}  \frac{1}{\epsilon_{r} }
\right] }, 
\\
(T^{-1})_{pq} &=&  
\frac{ -1 }
{\epsilon_{p}  \epsilon_{q} 
\left[
1 + \sum_{r=1}^{\ell}  \frac{1}{\epsilon_{r} }
\right] }
\hspace{0.5cm}
(p \neq q). 
\label{Tinv}
\end{eqnarray}
Then, the second term of $V^{-1}$ in (\ref{Vinv}) is given by 
\begin{eqnarray}
(V^{-1}_{(\text{2nd term})})_{i j} 
&=&
\frac{1}{\sigma^2_{u i}  \sigma^2_{u j}}
\left[
\sum_{p} H_{i p} (A^{-1})_{pp} H_{p j}  + 
\sum_{p \neq q} H_{i p} (A^{-1})_{pq} H_{q j} 
\right], 
\nonumber \\
&=&
\frac{1}{\sigma^2_{u i}  \sigma^2_{u j}}
\left[
\sum_{p}^{\ell} (A^{-1})_{pp} + 
\sum_{p \neq q}^{\ell} (A^{-1})_{pq} 
\right]. 
\label{Vinvkl}
\end{eqnarray}
We thus obtain $V^{-1}$ as 
\begin{eqnarray}
(V^{-1})_{i j} =
\frac{\delta_{i j}}{ \sigma^2_{u i}  } -
\frac{ 1 }
{\sigma^2_{u i}  \sigma^2_{u j} }
\frac{ \left(  \sum_{p=1}^{\ell}  \sigma^2_{c p} \right) }
{ \left[
1 + \left(
\sum_{k=1}^{n} \frac{1}{\sigma^2_{u k}}
\right) 
\left(  \sum_{p=1}^{\ell}  \sigma^2_{c p} \right) 
\right] }. 
\label{Vinv_final}
\end{eqnarray}

%%%%%%%%%%%%%%%% acknowledgments %%%%%%%%%%%%%
\begin{acknowledgments}
We thank Hiro Sugiyama and Osamu Yasuda for informative 
correspondences and helpful discussions on statistical procedure 
for our analysis. 
The encouraging comments from Raju Raghavan on the first version 
of the manuscript are gratefully acknowledged. 
One of the authors (H.M.) thanks 
Abdus Salam International Center for Theoretical Physics 
for hospitality where this work is completed. 
This work was supported in part by the Grant-in-Aid for Scientific Research, 
No. 16340078, Japan Society for the Promotion of Science.

\end{acknowledgments}
%%%%%%%%%%%%%%%%%%%%%%%%%%%%%%%%%%%%%%

\end{document}